\documentclass[aps,prx,twocolumn,superscriptaddress,showpacs,floatfix,longbibliography]{revtex4-2}
\usepackage{amsmath,amssymb,amsfonts,float,graphics,epsfig,epstopdf,color,verbatim,tabularx,bm,multirow,hyperref,ulem}
\hypersetup{
  colorlinks, linkcolor={blue},
  citecolor={blue}, urlcolor={blue}
}

\usepackage{wasysym}

\DeclareSymbolFont{sfletters}{OML}{cmbrm}{m}{it}
\DeclareMathSymbol{\sfeps}{\mathord}{sfletters}{"22}
\newcommand{\Tr}{\operatorname{Tr}}

\graphicspath{{Figures/}}
\newcommand{\gb}{\bar{g}}

\usepackage{times}

\usepackage[left]{lineno}
\begin{document}
%\linenumbers
\title{Monte Carlo study of the pseudogap and superconductivity \\ emerging from quantum magnetic fluctuations }

\author{Weilun Jiang}
\affiliation{Beijing National Laboratory for Condensed Matter Physics and Institute
of Physics, Chinese Academy of Sciences, Beijing 100190, China}
\affiliation{School of Physical Sciences, University of Chinese Academy of Sciences, Beĳing 100190, China}

\author{Yuzhi Liu}
\affiliation{Beijing National Laboratory for Condensed Matter Physics and Institute
	of Physics, Chinese Academy of Sciences, Beijing 100190, China}
\affiliation{School of Physical Sciences, University of Chinese Academy of Sciences, Beĳing 100190, China}

\author{Avraham Klein}
\affiliation{Department of Physics, Faculty of Natural Sciences, Ariel University, Ariel, Israel}

\author{Yuxuan Wang}
\affiliation{Department of Physics, University of Florida, Gainesville, FL 32601}

\author{Kai Sun}
\affiliation{Department of Physics, University of Michigan, Ann Arbor, MI 48109, USA}

\author{Andrey V. Chubukov}
\affiliation{School of Physics and Astronomy, University of Minnesota, Minneapolis, MN 55455, USA}

\author{Zi Yang Meng}
\email{zymeng@hku.hk}
\affiliation{Department of Physics and HKU-UCAS Joint Institute of Theoretical and Computational Physics, The University of Hong Kong, Pokfulam Road, Hong Kong SAR, China}
\affiliation{Beijing National Laboratory for Condensed Matter Physics and Institute of Physics, Chinese Academy of Sciences, Beijing 100190, China}

\begin{abstract}
\end{abstract}
\date{\today}
\maketitle

\noindent{\bf Abstract}\\

The origin of the pseudogap behavior,
found in many high-$T_\text{c}$ superconductors,
remains one
of the greatest puzzles in condensed matter physics. One possible mechanism
is
fermionic incoherence, which near a quantum critical point
allows pair formation but
suppresses
superconductivity.
Employing quantum Monte Carlo simulations
of  a model
of  itinerant fermions coupled to ferromagnetic spin fluctuations,
represented by a quantum rotor,
we report numerical evidence of pseudogap
behavior,
emerging from  pairing fluctuations in a quantum-critical non-Fermi liquid.
Specifically,
we observe enhanced pairing fluctuations and a partial gap opening in the fermionic spectrum.
However, the system remains non-superconducting until reaching a much lower temperature.
In the pseudogap regime
the system
displays a ``gap-filling" rather than ``gap-closing"
behavior,
similar to the
one observed in cuprate superconductors.
Our results
present direct evidence of the pseudogap state,
driven by superconducting fluctuations.

\vspace{\baselineskip}
\noindent{\bf Introduction}\\

Even though unconventional and high-$T_\text{c}$ superconductivity arises in a diverse set of materials, many of them share
similar
features
in their phase diagram. One prominent feature
is a superconducting (SC) dome, which emerges near the termination point of
a non-SC phase with either spin or charge order.
The second feature is
anomalous transport and non Fermi-liquid (nFL) behavior
around
the putative quantum critical points (QCP). These features have led to the proposal
that soft quantum-critical fluctuations of the order parameter serve as the
source for the universal behavior and mediate singular interaction that gives rise to
superconductivity with nontrivial pairing symmetry, strange metal behavior, and intertwined orders.

In many unconventional superconductors,
most notably the cuprates, there is a third
feature:
the ``pseudogap(PG)",
a gap-like
feature
in the fermionic spectrum above
the SC phase. Despite decades of investigation, the origin (or origins) of the PG remain intensely debated. One class of proposals names
exotic, possibly topological order in the particle-hole channel as the origin~\cite{Scheurer2018,Sachdev2018,Wu2018},
while another  points to
pairing fluctuations in the strong coupling regime ~\cite{Emery1995,Keimer2015,YMWu2020,Wang2020,Dahm2009,Reber2013,Kanigel2006}.
Substantial numerical efforts have been dedicated to the understanding of PG, see e.g. Refs. \cite{Gull2013,LeBlanc2015,Scheurer2018} and references therein.

\begin{figure*}[htp!]
	\includegraphics[width=\textwidth]{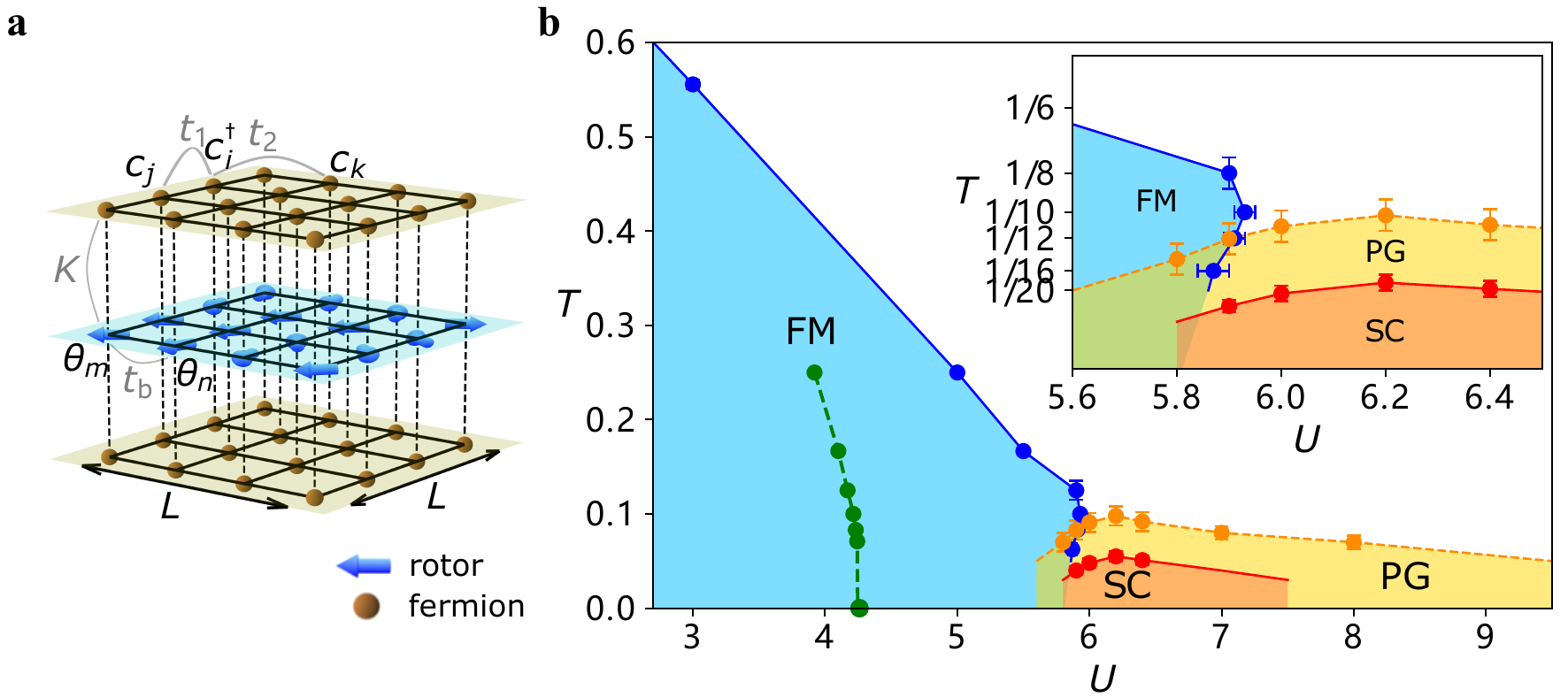}
	\caption{{\bf Model and Phase diagram.} {\bf a} Sketch of the model in Eq.~\eqref{eq:eq1}. Deep yellow dots and grids of the top and bottom layers represent fermion degrees of freedom with nearest hopping strength $t_1$ (e.g. $\hat c_i^{\dagger} \hat c_j$) and next-nearest hopping $t_2$ (e.g. $\hat c_i^{\dagger} \hat c_k$). Blue arrows and grid in the middle layers denote bosonic parts with an unit vector representing $\theta \in [0,2\pi)$ of the rotor on each site. The interaction between two rotors on nearest sites (e.g. $\theta_m$ and $\theta_n$) is $t_{\text{b}}$. The on-site coupling $K$ between fermions and bosons is shown by the vertical dashed lines. The system size is $L\times L$. {\bf b} $U-T$ phase diagram of the model obtained from QMC simulation. The inset
		zooms in to the
		vicinity of the pseudogap (PG, yellow),
		ferromagnetic (FM, blue) and superconducting (SC, orange) regions.
		The blue points
		on the
		FM
		phase boundary are determined by finite size scaling with fixed
		$T$ or $U$.
		Notably,
		for $U=5.9$, as temperature gets lower, the system first enters into
		the FM
		phase at
		$T\approx 0.13$, then exits it at
		$T\approx 0.08$.
		The yellow points of the PG boundary are determined from the onset of a PG in the single-particle spectrum,
		as shown in Fig.~\ref{fig:fig2}.
		The
		red points denoting an
		onset of s-wave superconductivity
		are determined from the onset of a full gap in the spectrum as well as Kosterlitz-Thouless scaling of the pairing susceptibility.
		The maximum of SC phase transition temperature $T_\text{c}$ is approximately 0.05. The green points and dashed line, are the phase boundary of the
		(uncoupled)
		quantum rotor model~\cite{WLJiang2019}.
                See the Supplementary Note 3 for additional details as well as a discussion of SC fluctuations above $T_\text{c}$. The errorbars of the points on the FM phase and SC phase boundaries are determined by the data collapse with fixed $T$ or $U$. For the PG phase boundary, the errorbars come from the uncertainty
                  % by indentying
                in identifying
                the onset of the minimum at $\omega=0$ for distinct temperatures of DOS.
              }
	\label{fig:fig1}
\end{figure*}

The understanding of the coupling between fermionic excitations
near the Fermi surface (FS) and bosonic quantum critical fluctuations ~\cite{Abanov2003,Metzner2003,Metlitski2010a,Metlitski2010b,Lee2018} is crucial
to describe these three features.
The development of quantum Monte Carlo (QMC) algorithms for a class of models of this type,
pioneered by Ref.
~\cite{Erez2012},
has
created a feasible
way
to study this physics in an unbiased manner
(see the reviews ~\cite{XYXu2019,Berg2019} and references within).
In
QMC models, FS fermions couple to bosonic fluctuations, representing certain collective modes of
low-energy fermions
~\cite{Schattner2016,Gerlach2017,Bauer2020,XYXuPRX2019,Liu2019,ChuangChen2020,ChuangChen2021,Gazit2020}.
The bosonic part is bestowed with independent (non-fermionic) dynamics and can be tuned to criticality
to mimic
the situation in real materials. Crucially, these models are free of the sign-problem plaguing most fermionic QMC, allowing for a realistic test of theory.

In this work, we investigate the PG physics via such a sign-free QMC simulation of fermions near a ferromagnetic QCP. We find robust signatures of a PG above the SC state and are able to study its spectral properties and its interplay with the dynamics of the ferromagnetic degrees of freedom. We also compare the numerical results with several theoretical predictions, and reconcile many key aspects of the two.

\vspace{\baselineskip}
\noindent{\bf Results}\\

\noindent{\bf Overview.} Before going into the details of our work, we present an overview of the essential features of our model and a summary of the main results.
  
% In this work, we consider
The model we choose to implement is a variant of a quantum critical
model, in which the bosons represent
critical ferromagnetic(FM) spin fluctuations (a ``spin-fermion'' model). When looking for a spectral property of the superconductivity, such as a PG, such a model has an advantage over analogous ones, e.g. antiferromagnetic or nematic models (see e.g. \cite{Berg2019}) because of the simplicity of the momentum structure of the FS (e.g. no hot or cold spots). Furthermore, compared to earlier
 sign-free QMC studies
on ferromagnetism, the coupling strength of our model is stronger in two
aspects.
First, the spin system is
an XY quantum rotor model
that is inherently more strongly fluctuating than
an Ising model, analyzed earlier~\cite{Xu2017,XYXu2020}. Second, the coupling constant $K$ between the fermionic and bosonic sectors is set to
larger values than in previous works. The larger coupling pushes the region of SC fluctuations up to  temperatures, where they are discernible in the numerical data. This in contrast with earlier works, where coupling strength was optimized to study normal state properties.
As we see below, the
larger coupling
% will
% allow
allows us to reveal
 the PG behavior.

In the normal state, at low enough temperatures we find in the bosonic sector near the QCP
an overdamped dynamics with linear frequency response ($z=2$ scaling). This is different from the $z=3$ behavior, found in Ising systems, and is a result of
a  non-conservation of the order parameter in our model.

In the
temperature range, where the bosonic susceptibility is linear in frequency, we observe several
remarkable features. The uniform susceptibility deviates from
Curie-Weiss behavior and actually becomes
weaker
at smaller  $T$.
In the fermionic sector, we find
a gap-like feature in
the density of states (DOS).
Unlike
in a BCS superconductor, the size of the gap remains roughly
independent on temperature, while the DOS
becomes
progressively depleted (filled) upon lowering (raising) temperature. Importantly,
the scaling behavior of the pairing susceptibility
clearly shows that the system
is not in a SC state. We thus identify the spectral gap in such a state as a PG.

We note that the ``gap-filling" behavior observed in our numerical results has also been observed in tunneling and photoemission experiments on the cuprates~\cite{Vishik2018},
and has been obtained
in a class of $\gamma-$models of quantum-critical pairing
~\cite{YMWu2020}.
Our results, obtained from unbiased large-scale QMC simulations,
confirm the existence of a PG behavior from pairing fluctuations in a quantum-critical system with itinerant fermions.

The quantum-critical spin dynamics and normal state fermionic properties that we find
are
consistent with recent theoretical predictions for nFLs at finite temperature , obtained
within the
modified Eliashberg theory~\cite{Chubukov2014,AKlein2020,XYXu2020}.
This allows
us to benchmark our simulations and extract relevant parameters from the observables (see Supplementary Note 4 for details).
The onset temperature for PG behavior,  $T_{\text{PG}}$, and SC $T_\text{c}$, extracted from QMC, are consistent with
theoretical predictions (see Ref.~\cite{YMWu2020} and Methods).
Our results therefore provide
an attempt to numerically realize the transition from nFL to PG and eventually to superconductivity, lending support to the scenario of pairing fluctuations driven PG phenomena.

Crucially, we view our finding of a PG as the evidence of a universal mechanism for the formation of a PG from SC fluctuations near a QCP,  not limited to the specific model of FM spin fluctuations that we used. We do not claim that we present a model for PG formation in a  specific material, but in view of our findings  we do expect SC fluctuations to be a contributing factor to PG formation in any system close enough to a QCP, independently of the specific origin of the pairing boson.

The PG, obtained in our work, comes from pairing fluctuations in a situation when the pairing is in turn  mediated by a propagator of a FM order parameter. While our model does not directly describe experimental situation in the cuprates, where antiferromagnetic fluctuations are often considered to be a pairing glue,  we argue that the mechanism for the PG formation, studied in our work, is a universal phenomenon of the pairing near a  quantum-critical point~\cite{YMWu2020}, and in this sense goes beyond  the specific model with FM flucutations.  We do expect the SC fluctuations to be a contributing factor to PG formation in any system close enough to a QCP, independently of the specific origin of the pairing glue.

\vspace{\baselineskip}
% \noindent{\bf Results}\\
\noindent{\bf Model.} We consider a model of
itinerant
fermions coupled to
SO(2) quantum rotors, as shown in Fig.~\ref{fig:fig1}{\bf a} (rotors are in the middle layer).
The model is described by
\begin{equation}
  \hat H = {{\hat H}_{\text{qr}}} + {{\hat H}_\text{f}}  +  {{\hat H}_\text{qr-f}},
  \label{eq:eq1}
\end{equation}
where
\begin{widetext}
\begin{eqnarray}
  {{\hat H}_{\text{qr}}} &=& \frac{U}{2}\sum\limits_i {\hat L_i^2}  - {t_\text{b}}\sum\limits_{\langle i,j \rangle } {\cos \left({{\hat \theta }_i} - {{\hat \theta }_j}\right)} \nonumber\\
  {{\hat H}_\text{f}} &=&  - t_1 \sum\limits_{\langle i,j \rangle \sigma \lambda } \hat c_{i\sigma \lambda }^ \dagger {{\hat c}_{j\sigma \lambda }} -t_2 \sum\limits_{\langle\langle i,j \rangle\rangle \sigma \lambda } \hat c_{i\sigma \lambda }^ \dagger {{\hat c}_{j\sigma \lambda }}  - \mu \sum\limits_{i\sigma \lambda } {{\hat n_{i\sigma \lambda }}} \nonumber\\
  {\hat H}_{\text{qr-f}} &=&
  - \frac{K}{2}\sum\limits_{i\lambda } {\left(\hat c_{i\lambda }^ \dagger {\sigma ^x}{{\hat c}_{i\lambda }} \cdot \cos {{\hat \theta }_i} + \hat c_{i\lambda }^ \dagger {\sigma ^y}{{\hat c}_{i\lambda }} \cdot \sin {{\hat \theta }_i}\right)}.
\end{eqnarray}
\end{widetext}

\begin{figure}[htp!]
	\includegraphics[width=\columnwidth]{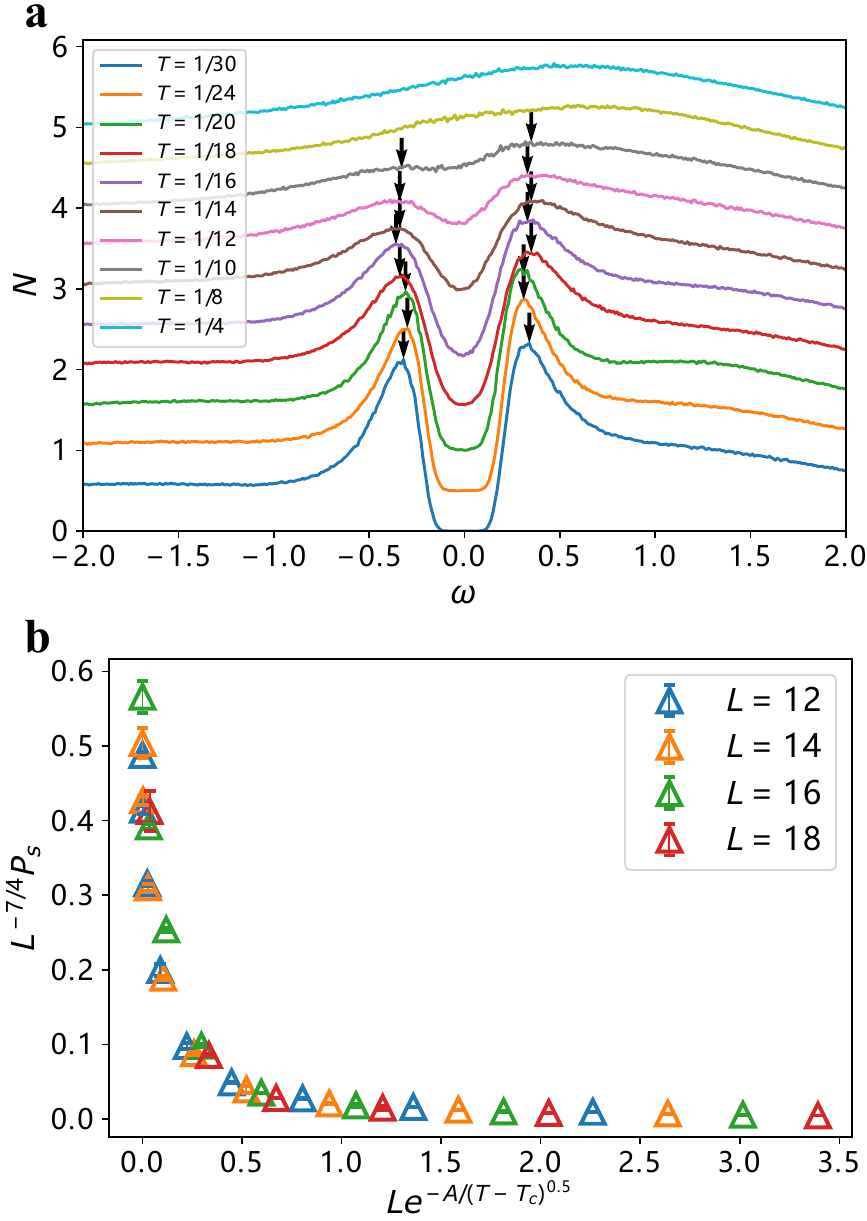}
	\caption{{\bf Pseudogap and superconductivity.} {\bf a}
  Local DOS $N(\omega)$ for various temperatures at $U=6$ with $L=12$. For
          $T=1/4$, far above
          the PG, the system
          exhibits a Fermi liquid spectrum.
          At $T_{\text{PG}}\approx 0.1$, the SC fluctuations begin to play important role and a
          noticable gap forms at $\omega=0$. This gap-forming temperature is consistent with the corresponding intermediate temperature scale in the dynamic bosonic susceptibility $\chi$ in Fig.~\ref{fig:fig4}{\bf b}. The gap
          minimum $N(\omega = 0)$ goes down with temperature and eventually reaches zero at $T \approx 0.05$, indicating the onset of a SC state as detected by the pairing susceptibility in panel b. At $T<0.05$, fully gapped spectrum corresponds to the behavior of the SC state.
          {\bf b} Data collapse of the pairing susceptibility $P_\text{s}$ versus temperature at $U=6$ for system sizes $L=12,14,16,18$ with statistical errors obtained by QMC simulations, consistent with a KT transition.
          The best fit coefficients are
          $A=0.75, T_\text{c}=0.048$, which is consistent with the temperature of the fully-gapped spectrum in {\bf a}.}
	\label{fig:fig2}
\end{figure}

The first
term
${{\hat H}_\text{qr}}$ describes a quantum rotor model on a square lattice. Here
$\hat L_i $
is the angular momentum
of 2D rotor $\hat \theta_i$ at site $i$.
The second term ${{\hat H}_\text{f}}$ describes two identical copies of spin-$1/2$ fermions on
a square lattice, with layer index $\lambda=1$ and $2$ representing the top and the bottom layers.
Fermions in each layer can hop between nearest-neighbor (next-nearest-neighbor) sites with hopping
amplitudes
$t_1$ ($t_2$), and  the chemical potential $\mu$
controls the fermion density. The last term ${\hat H}_{\text{qr-f}}$ couples
quantum rotors and fermions
via an on-site FM
interaction  that
tends to align XY component of a  fermion spin with the direction of a rotor on each site.

In  the absence of fermion-rotor coupling,
rotors
develop
quasi-long-range FM order
via a Kosterlitz-Thouless(KT) transition~\cite{Jose1977,WLJiang2019}. At zero temperature,
FM order becomes long range.
The KT transition line in $(T,U)$ plane
terminates at a QCP
at $(U/t_\text{b})_\text{c}=4.25(2)$
~\cite{Hasenbusch1999,WLJiang2019,ZYMeng2008}. As we turn on the fermion-rotor coupling, fermion contributions shift the KT phase boundary towards larger $U$ and $T$. More importantly, the
phase transition now involves fermion spins, which  at $T=0$ also order ferromagnetically below $U_\text{c}$.
This
allows us to study
quantum phenomena near a
FM QCP in a metal~\cite{Brando2016}.
Due to the anti-unitary symmetry and the presence of two copies of fermions, this model can be simulated via QMC techniques without the sign problem (see Supplementary Note 1 for details).
This setup then allows us to analyze  the universal behavior near a QCP
with
high
numerical accuracy and large system sizes.

We express all quantities in units of $t_\text{b}$.  In the  simulations we  set
$K=4$, $t_1=1$, $t_2=0.2$ and $\mu=0$.
We
varied $U$ and the temperature $T$ and
constructed the phase diagram of the model, Fig.~\ref{fig:fig1}{\bf b}, which
features a paramagnetic-ferromagnetic transition and several  other transitions/phases.
The magnetic transition at a finite temperature is of KT type.  As $U$ increases, the transition temperature decreases and  terminates at a QCP at $U_\text{c}$.   The $T=0$ transition upon varying $U$ belongs to XY universality class as the coupling to rotors creates an easy plane for fermion spins.
Fermion spins order ferromagnetically in the XY plane, breaking a spin-rotational symmetry.

\vspace{\baselineskip}
\noindent{\bf Pseudogap and superconductivity properties.}
We observe a
SC dome around the QCP.
Above the
dome, we find
evidence of PG behavior in the range of $T$,
whose width is
comparable to $T_\text{c}$.

First,
by measuring correlation functions of Cooper pairs in various
pairing channels, we found that the
dominant  pairing channel
is spin-triplet and odd under the interchange between the top and the  bottom layers (layer-singlet), i.e.,
$\Delta(\mathbf{r})=\frac{1}{\sqrt{2}} (\hat c_{\mathbf{r}1\uparrow} \hat c_{\mathbf{r}2\downarrow} - \hat c_{\mathbf{r}2\uparrow} \hat c_{\mathbf{r}1\downarrow}) = \frac{1}{\sqrt{2}} (\hat c_{\mathbf{r}1\uparrow} \hat c_{\mathbf{r}2\downarrow} + \hat c_{\mathbf{r}1\downarrow} \hat c_{\mathbf{r}2\uparrow})$, where $1$ and $2$ label layers.
In the classification of 2D irreducible representations, this is an $s$-wave gap, as $\Delta (0)$ is finite.
We verified (see Supplementary Figure 1, and 2) that the susceptibility in this channel  strongly increases when the system approaches a superconducting instability, while the susceptibilities in all other pairing channels remain small and do not increase.
This observation is a direct evidence
that
superconductivity
originates from the
interaction mediated by soft bosonic fluctuations,
associated with the QCP.
Indeed, it has long being known that near a
FM
quantum phase transition, soft dynamical bosonic fluctuations introduce an effective interaction that
is attractive in the spin-triplet channel. In the geometry of our model, there are two distinct types of spin-triplet pairing -- one is odd
under momentum inversion in a layer and even under layer interchange  (e.g., $p$-wave layer-triplet),  the other is even within each layer
and odd under layer interchange ($s$-wave layer-singlet).
By analogy with previous studies of the pairing mediated by small ${\bf q}$  fluctuations~\cite{Xu2017},
one expects the leading instability to be towards the $s$-wave layer-singlet, spin-triplet order.
The numerical
finding of the largest pairing correlations in this channel
thus affirms
the crucial role of soft
FM
bosonic fluctuations
in the formation of a SC dome.

\begin{figure}[t]
	\includegraphics[width=\columnwidth]{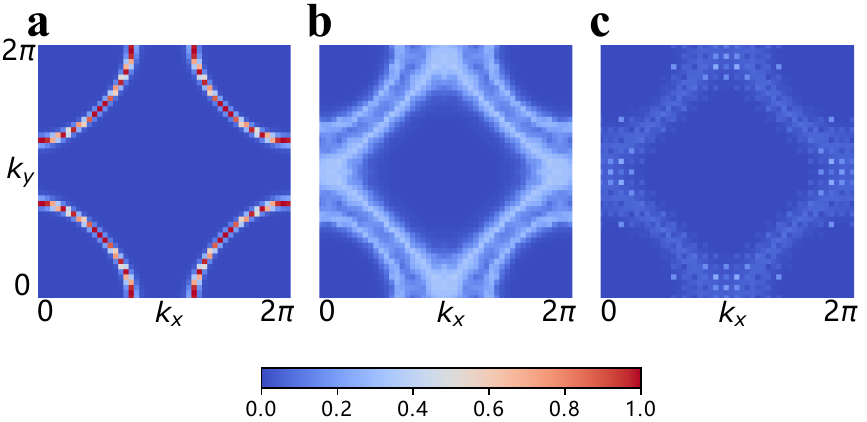}
	\caption{{\bf Re-entrance.} Evolution of the Fermi surface (FS) from non-interacting system with $H_\text{f}$ in {\bf a}, to the nFL FS subjected to strong FM correlation at $U=5.9$, $T=0.1$ in {\bf b}, and eventually to the FS in the PG phase at
          $U=5.9$,
          $T=0.05$ in {\bf c}. The spectral weights are normalized with the same scale. The system size is $L=12$ and the twisted boundary condition in the fermion hopping is applied such that the momentum resolution is 4 times larger in both $k_x$ and $k_y$ directions.
        }
	\label{fig:fig3}
\end{figure}

Second, we obtained
the fermionic spectral function, and then integrated it over $k-$space to obtain the
 DOS $N(\omega)$.
For this, we  first computed
the imaginary-time fermion Green's function
and then converted it to real frequency via stochastic analytic continuation (SAC) method (See Methods for details).  We show the results for $N(\omega)$ in
Fig.~\ref{fig:fig2}{\bf a}. At
low $T$, inside the
SC dome, there is  clear evidence for an
$s$-wave gap.
The data shows that, that as $T$ increases,  the magnitude of the gap slightly increases, rather than shrinks, as would be the case in a
BCS superconductor. Simultaneously,
$N(\omega)$ for $\omega$ smaller than the gap increases and
gradually
fills in the states within
the gap,
ultimately restoring its normal-state value.   This phenomenon has been termed
gap-filling. It is
qualitatively in agreement
with experimental observations in many strongly-correlated unconventional superconductors at $T \geq T_\text{c}$~\cite{Kanigel2006,Reber2012,Reber2013,Keimer2015}, At smaller $T \leq T_\text{c}$,  the DOS displays gap-closing
behavior, like in a conventional BCS superconductor.
Guided by the
experimental evidence~\cite{Kanigel2006,Reber2013}  that gap-filling behavior holds at $T \geq T_\text{c}$,
we defined the PG region as the one where
the
DOS gets filled in upon increasing $T$.
We set the lower
boundary of this region to
where the DOS at the Fermi energy
significantly deviates from  thermally activated behavior of $e^{-\Delta/k_\text{B} T}$.
The upper
boundary of the PG region is set at $T_{\text{PG}}$, at which
the dip of $N(\omega)$ at the Fermi energy becomes invisible.
The PG region, obtained this way, is
plotted in yellow
in Fig.~\ref{fig:fig1}{\bf b}.

Third,  to  determine the actual SC
transition temperature, $T_\text{c}$,
we performed scaling analysis
of
the pairing susceptibility $P_\text{s}= \frac{1}{L^2}\int_0^{\beta} \sum_i (\Delta^{\dagger}(\mathbf{r}_i,\tau) \Delta(\mathbf{0},0))$,
using
KT scaling
for the pairing susceptibility
$P_\text{s} = L^{2-\eta} f(L\cdot \exp(-\frac{A}{(T-T_\text{c})^{1/2}}))$ for $T > T_\text{c}$
with $\eta_{\text{KT}}=1/4$~\cite{Paiva2004,costa2018phonon,ChuangChen2021}. We show the results in  Fig.~\ref{fig:fig2}{\bf b}. The
data  for $P_\text{s}$ for various system sizes and temperatures collapse onto a single curve. We fitted the curve by the formula above and extracted
$T_\text{c}=0.048$. This agrees
with the lower boundary of the
PG region. The upper boundary, $T_{\text{PG}}$, is about twice larger in our simulations, $T_{\text{PG}} \sim 0.1$.
We
also computed
the superfluid density, $\rho_\text{s}(T)$,
which has been widely used
to estimate
$T_\text{c}$
in QMC simulations.
This is done by
detecting
the temperature $T_\rho$
at which
$\rho_\text{s}(T_\rho)= \alpha T_\rho$, where $\alpha$ is a
dimensionless constant~\cite{costa2018phonon},
usually set to $2/\pi$, based on the analysis of the XY model~\cite{Pokrovsky_1979}.
This criterion, although qualitatively correct, typically overestimates $T_\text{c}$~\cite{Paiva2004}. In our
case,
we found $T_\text{c} < T_\rho \sim T_{\text{PG}}$.
We discuss our analysis of $\rho_{\text{s}}$ in some length in Supplementary Note 3.

\begin{figure*}[htp!]
\includegraphics[width=\textwidth]{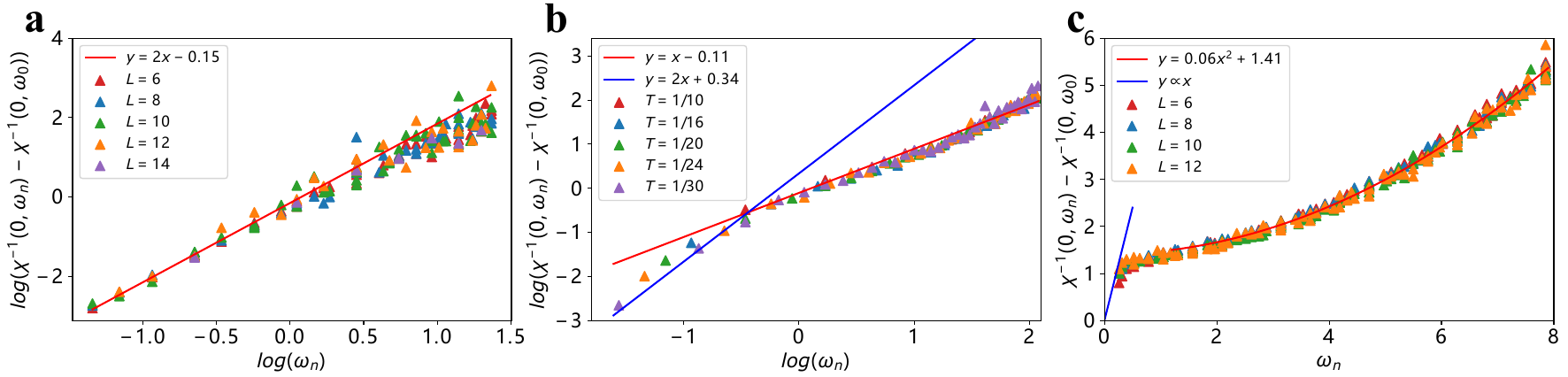}
\caption{{\bf Magnetic dynamics.} Inverse bosonic dynamic susceptibility $\chi$ versus $\omega_n$ in three different regions, at $U=3, 6, 8$, corresponding {\bf a} in the FM phase, {\bf b} in PG and SC phases, and {\bf c} disordered phase. {\bf a} log-log plot for various system size $L=6,8,10,12,14$, each of which includes various $\beta=12,16,20,24$. Red line is a quadratic line of $\chi^{-1} \sim \omega^{2}$ for low frequency part $\omega_n<1$. {\bf b} log-log plot for various $\beta=10, 16, 20, 24, 30$ with $L=12$. At temperature $T=0.1$ ($\beta=10$), the fermions are in the quantum critical regime, the bosonic susceptibility is the linear function of $\omega$, as indicated by the orange line. When temperature gets lower, the fermion goes into PG phase, prompting the bosonic scaling behavior to deriviate from linear function. And upon entering the SC phase, the $\chi^{-1} \sim \omega^2$ (the blue line, a guide to the eye) as in the FM phase. {\bf c} Bosonic susceptibility in the disordered phase at $U=8$ plotted for system size $L=6,8,10,12$ with various $\beta=12,16,20,24$. At high frequency all data points successfully merge together, as indicated by the red line quadratic in $\omega$ for $\omega>1$. At low frequency, the $\chi^{-1}\sim \omega$ as indicated by the blue line, which is a guide to the eye, due to the non-conserved rotor order parameter.}
\label{fig:fig4}
\end{figure*}

We  analyzed the QMC data within the quantum critical theory of itinerant ferromagnets~\cite{Rech2006,Maslov2009},
extended to finite $T$~\cite{AKlein2020} and modified to include two layers of itinerant fermions and superconductivity.
We computed
fermionic and bosonic self-energies near $U_\text{c}$ and found good agreement with the simulations in the normal state (see Supplementary Note 4).
We  extracted
the effective fermion-boson coupling from this comparison, and used it  to compute the onset temperature for the pairing within the Eliashberg theory for quantum-critical pairing~\cite{YMWu2020}.  This theory does not differentiate between pair formation and superconductivity, hence the result  has to be compared with $T_{\text{PG}}$, extracted from simulations.
We obtained theoretical $T_{\text{PG}} \sim 0.08$,  quite consistent with $T_{\text{PG}} \sim 0.1$, extracted from QMC data, see Fig.\ref{fig:fig1}{\bf b}. Further, Eliashberg calculations below $T_{\text{PG}}$ show gap-closing behavior at small $T$ and gap-filling behavior at $T \leq T_{\text{PG}}$. The boundary between two regimes  has been associated with the actual $T_\text{c}$, based on the analysis of phase fluctuations~\cite{YMWu2020}. 
We show this
in
the phase diagram in Fig.\ref{fig:fig1}{\bf b}.
Based on this comparison, we argue that
our unbiased numerical  QMC simulations are consistent with the theory and
provide strong
evidence for
PG behavior, originating from  preformed pairs above $T_\text{c}$,  near a FM
QCP  in a metal.

We note that previous QMC work (see e.g. \cite{Schattner2016,Gerlach2017} for an antiferromagnetic model) found an SC dome surrounded by a region of SC fluctuations. These were determined by comparing $T_\rho$ determined from the BKT criterion for $\rho_\text{s}$ (see above), and the temperature $T_{\text{dia}}$ at which the system showed diamagnetic behavior, as evidenced by a sign-change of the appropriate current-current correlator. While we too find such fluctuations (see Supplementary Note 3). We stress that the region of gap-filling behavior is predominantly at
$T_\text{c} < T < T_{\text{PG}}\sim T_\rho$, and is therefore distinct from fluctuations on the scale of $T_{\text{dia}}$. Indeed, from our numerics we observe that 
$T_{\text{PG}} < T_{\text{dia}}$. We emphasize that in the thermodynamic limit, $T_{\text{PG}}$ and $T_{\text{dia}}$ do not correspond to phase transitions, but rather mark crossover regions.

\vspace{\baselineskip}
\noindent{\bf
  Magnetic dynamics and re-entrance effect.} The
pairing behavior also has an
impact on the magnetic phase transition and the quantum dynamics of the rotors.
As shown in
Fig.~\ref{fig:fig1}{\bf b}, the phase boundary of the
paramagnetic-ferromagnetic transition exhibits a re-entrance behavior at $U\sim 5.9$, close to the QCP.
For example, at $U=5.9$, upon reducing the temperature, the system first enters the FM state and then returns to the paramagnetic one, i.e. there is a ``back-bending'' of the transition line to the FM state.
This can also be seen from the Fermi surface behavior. In Fig.~\ref{fig:fig3}, we plot the Fermi surface, $G(\mathbf{k},\tau=\beta/2) \sim N(\mathbf{k},\omega=0)$, evolution with temperature.
At intermediate temperature $T=0.1$, the Fermi surface splits due to the ferromangetic order. However,
the split vanishes both either increasing or lowering the temperature.
We believe that the
re-entrance phenomenon is
a
consequence of the PG and SC fluctuations, which
suppress the fermion DOS and hence the electron-hole contribution to magnetic order \cite{Moon2009,Moon2010}. Similar behavior has been seen previously in an antiferromagnetic model~\cite{Schattner2016}, but no PG was reported there.
We emphasize that the paramagnetic-ferromagnetic  phase boundary starts to bend to the left roughly at $T_{\text{PG}}$, which is
well above the SC
dome,
indicating that SC fluctuations
without phase coherence in the PG region are responsible for the magnetic dynamics.

We note, that
in the absence of an SC dome,
previous works
(see e.g. \cite{Belitz1997,Maslov2009,Brando2016,Green2018}) have shown that itinerant FM QCPs are unstable to a first-order transition driven by normal state magnetic fluctuations, which can cut off the FM phase at larger $U$'s, similar to the behavior seen in our simulations.
In our results, we do not observe clear evidence of a first-order magnetic transition, and
the correlation of the back-bending with $T_{\text{PG}}$
implies that
for the parameters that we used the physics is driven chiefly by
SC rather than magnetic fluctuations.

In addition, we measured the
inverse dynamical bosonic susceptibility of the rotors across different regions of the phase diagram.
Our results are summarized in Fig.~\ref{fig:fig4}, showing data for three representative $U$ at various temperatures. To study the dynamics, we subtract the static part of the inverse susceptibility and focus on the spin polarization  $\chi^{-1}(\mathbf{q},\omega) - \chi^{-1}(\mathbf{q}=0,\omega=0)$ (for details of what follows see Supplementary Note 2 and 3). Deep in the FM phase, Fig.~\ref{fig:fig4}{\bf a}, we find an $\omega^2$ dependence (dynamical exponent $z=1$). This is similar to that of the bare rotor model, and indicates that the fermionic contribution to the dynamics is negligible because of the spin gap. Similarly, deep in the Fermi liquid phase, Fig.~\ref{fig:fig4}{\bf c}, we find an $\omega^2$ dependence, except at the lowest frequencies, which furthermore extrapolates to a nonzero value. The saturation is readily understood as resulting from the non-analyticity of the Lindhard function which implies $\chi^{-1}(\mathbf{q}=0,\omega\to 0) \neq \chi^{-1}(\mathbf{q}\to0,\omega=0)$ at weak coupling. In the quantum-critical regime, Fig.~\ref{fig:fig4}{\bf b}, we find a qualitatively different behavior indicating strong fermionic correlations.

First, at higher frequencies we find a linear frequency dependence ($z=2$), which does not saturate to a finite value. This is surprising at first glance, since Landau damping for a ferromagnet has an $\omega/q$ form ($z=3$) rather than linear $\omega$. We note that, in purely electronic models $\chi^{-1}(\mathbf{q},\omega)$ is required to be non-analytic at any coupling strength due to spin conservation, and non-analytic behavior was seen previously in simulations of Ising-ferromagnets. However, in our simulations of the XY model, the order parameter is not conserved, leading to linear frequency dependence, in direct contrast to Ising model studied in Refs.~\cite{Xu2017,XYXu2020}.

Second, at lower Matsubara frequencies accessible  at lower temperatures, the $\omega^2$ behavior is again restored even in the quantum-critical region. As discussed above, this is again a direct result of the formation of a gap - this time the pseudogap, which depletes the low-energy fermion density of states and reduces the fermionic feedback  on the bosons.

From the analysis above, we see that the spin dynamics is consistent with the quantum-critical behavior and PG physics.

\vspace{\baselineskip}
\noindent {\bf Discussion}\\

In this work, we performed a large-scale quantum Monte Carlo simulation of a FM spin-fermion model.
We reported direct spectral and thermodynamic evidence of the formation of a PG prior to the SC transition. Within such a PG phase, the temperature evolution of the fermion spectral gap exhibits a gap-filling behavior, in sharp contrast with that of a conventional superconductor.
Moreover,
we found that the dynamics of the spin fluctuations display a different behavior than the well-known Landau damping behavior with $z=3$.

Remarkably, we were able to reconcile all these features with theoretical predictions of Eliashberg theory and its generalization to the $\gamma$-model.
Experimentally,  PG phases have been observed in various unconventional superconductors \cite{Kasahara2016,Oh2021}, most notably the cuprates \cite{RevModPhys.78.17}. Our results imply that a PG arising from strong dynamical fluctuations should be
ubiquitous in quantum-critical metals, and we expect this to be a
fruitful direction for future research.

\vspace{\baselineskip}
\noindent {\bf Methods}\\

\noindent {\bf QMC simulations and data analysis.} We employ the determinant quantum Monte Carlo (DQMC) method~\cite{Xu2017,XYXu2019} to simulate the Hamiltonian in Eq.~\eqref{eq:eq1}. The quantum rotor model plays the role of the auxiliary field in the conventional DQMC and the quantum rotor model can be efficiently simulated with non-local update scheme developed in our previous work~\cite{WLJiang2019}. For each realization of the rotor in space-time, the fermion determinant is evaluated with the kinetic part and the coupling part of the Hamiltonian included as the configurational weight and the Markov chain of the Monte Carlo process is carried out according the weight. Detailed measurements of the physical observables are given in the Supplementary Note 2.

In order to obtain real-frequency spectral functions, the SAC scheme is employed to obtain the spectral function $N(\omega)$ from the imaginary-time correlation function $G(\tau)$,
\begin{equation}
G(\tau)=\int_{-\infty}^{\infty} d \omega \frac{e^{-\omega(\tau-\beta / 2)}}{2 \cosh (\beta \omega / 2)} N(\omega)
\end{equation}
It is known that  the problem of inverting the Laplace transform is equivalent to find the most probable spectra $N(\omega)$ out of its exponentially many suggestions to match the QMC correlation function $G(\tau)$ with respect to its stochastic errors, and such transformation has been converted to a Monte Carlo sampling process~\cite{Sandvik1998,Beach2004,Sandvik2016}.
This QMC-SAC approach has been successfully applied to quantum magnets and interacting fermion systems ranging from the simple square lattice Heisenberg antiferromagnet~\cite{Shao2017} to deconfined quantum critical point and quantum spin liquids with their fractionalized excitations~\cite{GYSun2018,Ma2018,CKZhou2020} and to the continuum model of twisted bilayer graphene and benchmarked with the exact solution at the chiral limit~\cite{XuZhang2021,GPPan2021}.

\vspace{\baselineskip}
\noindent {\bf Theoretical analysis.} We analyzed the QMC data for fermionic and bosonic response using the modified Eliashberg theory, which is a low energy effective  dynamical theory for itinerant fermions near a QCP at finite temperatures. The theory accepts as parameters the static properties of a coupled fermion-boson system near a QCP, e.q. fermion bandstructure, bosonic susceptibility, etc., and computes the dynamical response of the system in terms of the fermionic self energy $\Sigma(\mathbf{k},\omega_n)$ and bosonic self energy (polarization) $\Pi(\mathbf{q}, \Omega_n)$, taking into account the low energy excitations near the FS. It accounts for deviations from the canonical $T\to 0$ quantum critical behavior, e.g. deviations from the  $\Sigma \sim \omega_n^{2/3}$ nFL self energy, and from the Landau damping $\Pi \sim \Omega_n/(v_F |\mathbf q|)$ as discussed in the main text. For details on the method see Refs.~\cite{AKlein2020,XYXu2020}.

We applied the theory to our QMC data, both to verify our assumptions on the normal state of the system and to extract the effective fermion-boson coupling. In the bare theory, the coupling $\gb \sim K^2$, but it is renormalized by fermions with energies of order of the bandwidth, so it should be extracted by fitting from the QMC data. We present results for $U=6$ which is almost above the QCP in Supplementary Figure 13, showing good agreement between theory and data. For details of the fitting procedure and a discussion of the quality of the fits are presented in Supplementary Note 4. We found $\gb = 6.3 \pm 0.2$, representing about a $20\%$ renormalization of the bare vertex $K$, which is consistent with earlier works~\cite{AKlein2020}.

Finally, we used the obtained $\gb$ to predict $T_{\text{PG}}$ within Eliashberg theory (the $\gamma-$model). Our model corresponds to $\gamma = 1/3$. The analytical prediction for $T_{\text{PG}}$ can be found in Ref.~\cite{YMWu2020}, and details of the conversion from our $\gb$ to the $\gamma-$model parameters are in the Supplementary Note 4. We found $T_{\text{\text{PG}}} \approx 0.08$, in good agreement with the QMC $T_{\text{PG}} \sim 0.1$.

\vspace{\baselineskip}
\noindent {\bf Data availability} \\
The data that support the findings of this study are available from the corresponding author upon reasonable request.

\vspace{\baselineskip}
\noindent {\bf Code availability} \\
All numerical codes in this paper are available upon reasonable request to the authors.

\vspace{\baselineskip}
\noindent {\bf Acknowledgements} \\
We thank R.M. Fernandes, M.H. Christensen, Y. Schattner, E. Berg, and X. Wang for valuable discussions. WLJ thanks Z. Liu for the support of the code, and G. Pan for the helpful suggestions.
WLJ, YZL and ZYM acknowledge support from the RGC of Hong Kong SAR of China (Grant Nos. 17303019,
17301420, 17301721 and AoE/P-701/20), the Strategic
Priority Research Program of the Chinese Academy of
Sciences (Grant No. XDB33000000), the K. C. Wong
Education Foundation (Grant No. GJTD-2020-01) and
the Seed Funding "Quantum-Inspired explainable-AI" at
the HKU-TCL Joint Research Centre for Artificial Intelligence. We thank the Center for Quantum Simulation Sciences in the Institute of Physics, Chinese Academy of Sciences, the Computational Initiative at the Faculty of Science and the Information Technology Services at the University of Hong Kong and the Tianhe platforms at the National Supercomputer Centers in Tianjin
and Guangzhou for their technical support and generous allocation of CPU time. The authors also acknowledge Beijng PARATERA Tech CO.,Ltd.(https://www.
paratera.com/) for providing HPC resources that have
contributed to the research results reported within this
paper. The work by AVC was supported by  the Office of Basic Energy Sciences, U.S. Department of Energy, under award  DE-SC0014402. AK and AVC acknowledge the hospitality of KITP at UCSB, where part of the work has been conducted. The research at KITP is supported by the National Science Foundation under Grant No. NSF PHY-1748958. YW is supported by startup funds at the University of Florida and by NSF under award number DMR-2045871.

\vspace{\baselineskip}
\noindent {\bf Author contributions}\\
AK, YW, KS, AVC and ZYM initiated the work. WLJ and YZL developed the program and performed the QMC calculations, WLJ, YZL and ZYM carried out the numerical data analysis. AK, YW, KS and AVC performed the theory analysis. All authors wrote the manuscript together.

%\vspace{\baselineskip}
%\noindent {\bf Competing interests}\\
%The authors declare no competing interests.

\clearpage
\begin{widetext}
  %\linenumbers
  \begin{center}
  \noindent {\bf \Large Supplementary Information for : Monte Carlo study of the pseudogap and superconductivity emerging from quantum magnetic fluctuations}
  \end{center}
  
  \vskip3mm
  
  In this supplementary information, we present the details of quantum Monte Carlo (QMC) implementation and more results of the phase diagram in different phases, as well as the theoretical analysis upon the QMC data.
  
  \vspace{\baselineskip}
  \noindent {\bf Supplementary Note 1: Details of QMC simulation.}\\
  \vskip1mm
  \noindent{\bf A. Quantum Rotor Model.}
  \label{app:appA}
  
  The Monte Carlo simulation on the quantum rotor model (QRM) starts from employing a proper basis for the Hamiltonian. As shown in Eq. 2 in main text, the boson part has global U(1) symmetry under $\theta$ representation. We adopt the representation of the angle variable $\theta$ for each site, ranging between $[0,2\pi)$, which are the eigenstates of the potential part. Using the canonical commutation relation $[ \hat \theta_i , \hat n_j ]= i\delta_{i,j} $, the QRM Hamiltonian can be expressed as,
  \begin{eqnarray}
     H_{\text{qr}} &=& \hat T+\hat U \\
     &=& \frac{U}{2}\sum_i \big(-i\frac{\partial}{\partial \hat \theta _i}\big)^2  - t_{\text{b}}\sum_{ \langle i,j \rangle } \cos (\hat \theta _i - \hat \theta_j)
  \end{eqnarray}
  and the partition function obeys,
  \begin{equation}
  Z = \text{Tr} \left\{ e^{-\beta[-\frac{U}{2}\sum_i\frac{\partial^2}{\partial \hat \theta^{2}_{i}}-t_{\text{b}}\sum_{\langle i,j \rangle}\cos(\hat \theta_i - \hat \theta_j)]} \right\}
  \end{equation}
  Using Trotter decomposition, we can divided $\beta$ into $M$ slices with step $\Delta \tau=\beta / M$, and insert the complete sets of the $\{ \theta_i \}$ at each time slice. We have,
  \begin{equation}
     Z = \int \mathcal{D}\theta \prod_{l=0}^{M-1} \langle\{\theta(l+1)\}| e^{-\Delta\tau \hat T} e^{-\Delta\tau \hat V}|\{\theta(l)\}\rangle
  \end{equation}
  The states follow the periodic boundary condition $\{\theta(M)\}=\{\theta(0)\}$. For the potential part, $\theta_i(l)$-s are the eigenstates of $V$ and can be directly calculated. For the kinetic part, if one inserts a complete basis of $J_i(l)$ as the integer-valued angular momentum at site $i$ and time slice $l$,
  % the left kinetic part writes,
  one finds that 
  \begin{equation}
     T(l) = \sum_{\{J\}}\prod_{i} e^{-\frac{\Delta\tau U}{2}[J_i(l)]^2} \langle \theta_i(l+1)|J_i(l)\rangle \langle J_i(l)|\theta_i(l)\rangle,
  \end{equation}
  The term $\langle \theta_i(l) | J_i(l) \rangle$ has a complex value $e^{i J_i(l) \theta_i(l)}$. Next, we transfer the square term of $J_i(l)$ into linear term with the help of the Poisson summation formula,
  \begin{eqnarray}
  T(l) &=& \prod_{i} \sum_{J}e^{-\frac{\Delta\tau U}{2} J^2}e^{iJ(\theta_i(l)-\theta_i(l+1))} \nonumber\\
  &=& \prod_{i} \sum_{m=-\infty}^{\infty}\int_{-\infty}^{\infty}dJ e^{2\pi i J m}e^{-\frac{\Delta\tau U}{2}J^2}e^{iJ(\theta_i(l)-\theta_i(l+1))} \nonumber\\
  &=& \prod_{i} \sum_{m=-\infty}^{\infty}\sqrt{\frac{2\pi}{\Delta\tau U}} e^{-\frac{1}{2\Delta\tau U}(\theta_i(l)-\theta_i(l+1) - 2\pi m)^2}.
  \end{eqnarray}
  Then, we modify this  by a Villain approximation to
  \begin{equation}
  T(l) \approx \prod_{i} e^{\frac{1}{\Delta \tau U}\cos(\theta_i(l)-\theta_i(l+1))}
  \label{eq:eq6}
  \end{equation}
  where the kinetic part of QRM can be regarded as the effective interaction along imaginary time axis. We finally map the QRM to 3D anisotropic XY model~\cite{WLJiang2019}, and the space-time configuration of the rotors, as shown in the Fig.\ 1a of the main text, plays the role of the usual auxiliary field for the determinant QMC simulations, which we will discuss next. 
  
  \vskip1mm
  \noindent{\bf B. Determinantal QMC implementations.}
  \label{app:appB}
  
  The determinantal quantum Monte Carlo (DQMC) is designed to deal with the interacting fermion lattice model with quartic interactions~\cite{BSS1981} and to decouple the quartic interactions into auxiliary bosonic fields coupled with fermion bilinears. In recent years, there new developments allow one to bestow the bosonic auxiliary field with a bosonic Hamiltonian and investigate the situation of the critical bosons coupled with various Fermi surface geometries~\cite{XYXu2019}, which is the path we take in this work. 
  
  In DQMC, one transfers the non-eigenstates to a series of classical configurations, such as the space-time rotor configurations in previous section, and then samples in the resulting configuration space.
  % in the form matrix expressions (or the determinant of matrices).
  To start with, one writes down the path integral of partition function,
  \begin{equation}
     Z=\Tr\{e^{-\beta \hat H}\}=\Tr\{\prod_{m=1}^M e^{-\Delta\tau \hat H}\}
  \end{equation}
  Here, $M=\beta / d\tau$, denoting the number of the imaginary time slices. $\hat H$ is the total Hamiltonian and contains both bosonic and fermionic parts, and their interaction. The trace operation can be divided into trace for fermions $\Tr_{\text{F}}$ and bosons $\int d\theta_i$, where we express the bosonic degrees of freedom as $\theta$ for each site. Next, we insert a series of unit operators with periodic boundary conditions $\{\theta(M)\}=\{\theta(0)\}$ and make $\Delta \tau \rightarrow 0$,
  \begin{eqnarray}
     Z&=&\Tr_{\text{F}}\left\{\int \mathcal{D}\theta \langle\{\theta\} | \prod_{l=0}^{M-1} e^{-\Delta \tau \hat H} | \{\theta\} \rangle\right\} \\
      &=&\Tr_{\text{F}}\left\{\int \mathcal{D}\theta \prod_{l=0}^{M-1} \langle\{\theta(l+1)\} | e^{-\Delta \tau \hat H_{\text{qr}}} e^{-\Delta \tau \hat H_{\text{F}}} e^{-\Delta \tau \hat H_{\text{qr}-{\text{f}}}} | \{\theta(l)\} \rangle\right\}
  \end{eqnarray}
  Next, utilizing the Supplementary Equation ~\eqref{eq:eq6}, we write the partition function as 
  \begin{equation}
     Z=\int \mathcal{D}\theta \left(\prod_{l=0}^{M-1} \prod_{i} e^{\frac{1}{\Delta \tau U}\cos(\theta_i(l)-\theta_i(l+1))} \right) \left(\prod_{l=0}^{M-1} e^{\Delta \tau t_{b} \sum_{\langle i, j\rangle} \cos (\theta_{i}(l)-\theta_{j}(l))}\right) \Tr_{\text{F}} \left\{\prod_{l=0}^{M-1} e^{-\Delta \tau \hat H_{\text{F}}} e^{-\Delta \tau \hat H_{\text{qr}-{\text{f}}}} \right\}
  \end{equation}
  The bosonic part can be taken out of the fermion trace. Furthermore, the kinetic part of free fermion is independent of configuration and can be calculated at the beginning of the simulation, while the interaction part of boson and fermion depends on the configurations of ${\theta}$. The calculation process on $\text{Tr}_{\text{F}}$ is always displayed as the determinant. %???
  Finally, the total weight of configuration is,
  \begin{eqnarray}
     Z&=&\int \mathcal{D}\theta \ \text{W}_{\text{b}}(\{ \theta \}) \det(\mathbf{1}+ \prod_{l=0}^{M-1} e^{-\Delta \tau H_{\text{F}}} e^{-\Delta \tau H_{\text{qr}-{\text{f}} {\{ \theta(l) \}}}}) \nonumber\\
     &=& \int \mathcal{D}\theta \ \text{W}_{\text{b}}(\{ \theta \}) \det(\mathbf{1}+B(\beta,0)_{\{ \theta \}}) \nonumber\\
     &=& \int \mathcal{D}\theta \ \text{W}_{\text{b}}(\{ \theta \}) \text{W}_{\text{F}}(\{ \theta \})
  \end{eqnarray}  
  where $\text{W}_{\text{b}}(\{ \theta \})$ is the weight of bosonic part, and $H_{\text{F}}, H_{\text{qr}-{\text{f}}}$ is the matrix in fermionic layer, spin, coordinate representation. So far, we have mapped the model to a series of classical configurations and obtained its weight. Using Markov chain, we sample the configuration of $\{\theta\}$ and implement both local update schemes and global updates - a Wolff update scheme - to avoid critical slowing down, see algorithm analysis and details of QRM in~\cite{WLJiang2019}.
  
  \vskip1mm
  \noindent{\bf C. QMC sign problem.}
  \label{app:appC}
  
  We find an antiunitray transformation $\mathcal{K}=i \sigma_y K$ under which the model is invariant, where $\sigma_y$ is a Pauli matrix in the layer basis, and $K$ is the complex conjugation operator. Thus the model is free of the sign problem, and
  \begin{equation}
     \text{W}_{\text{F}}= \det(\mathbf{1}+B'(\beta,0)_{\{ \theta \}})^2
  \end{equation}
  $B'$ is $2N \times 2N$ dimension matrix for single layer fermions, where $N=L \times L$ is the number of sites. Since the model is symmetric for two layers, the Green's function is the same for both layers with same site and spin index, i.e. $G_{11}^{\sigma \sigma'}=G_{22}^{\sigma \sigma'}$, where $G_{\lambda \lambda'}^{\sigma \sigma'}=\langle T \hat c_{i \sigma \lambda} \hat c_{j \sigma \lambda'}^{\dagger} \rangle$.
  
  \vskip1mm
  \noindent{\bf D. Controlling finite-size effects.}
  \label{app:appD}
  
  The simulation of DQMC is restricted to  finite system sizes, and its computational complexity scales as $O(\beta N^{3})$~\cite{XYXu2019}. Therefore, it is necessary to reduce the finite size effect in simulations as much as possible.
  % to save computing time and resource. %For the free electron system, the small system has fewer eigenstates, thus the energy level and DOS is more discontinuous than larger system. Therefore, the coupled mode with bosonic part is hard to compare with that of the infinite system. 
  To increase the momentum resolution on finite size simulations, we introduce a magnetic field perpendicular to the lattice plane called $z$-direction flux. The magnetic field changes the dispersion relation of the free system to mimic the DOS of the infinite system~\cite{Assaad2002}. Since the lattice site is finite, the flux must be quantized. The magnetic field is introduced via the Peirls phase factors on the bonds,
  \begin{equation}
     \hat c_{i\sigma \lambda }^ {\dagger} {{\hat c}_{j\sigma \lambda }} \rightarrow e^{i\int_{\mathbf{r}_i}^{\mathbf{r}_j} \mathbf{A}_{\sigma \lambda}(\mathbf{r}) d\mathbf{r}} \hat c_{i\sigma \lambda }^ \dagger {{\hat c}_{j\sigma \lambda }} = e^{iA_{ij}} \hat c_{i\sigma \lambda }^ \dagger {{\hat c}_{j\sigma \lambda }} 
  \end{equation}   
  with $\mathbf{B}=\triangledown \times \mathbf{A}$ and $\Phi_0$ the flux quanta. We take the Landau gauge $\mathbf{A}(\mathbf{r})=-B(y,0,0)$, which is independent of spin and layer index. To satisfy the periodic boundary condition, the boundary hopping terms must have different form compared with that of the inner bonds. Furthermore, we hope %???
  the flux on each area of lattice plane is equivalent. Since the model has the next-nearest hopping term, the square area encircled by four adjacent sites can be divided into four triangular parts. We design the magnetic field to satisfy this condition and for the nearest-neighbor hopping the phases $A_{ij}$ read,
  \begin{equation}
     A_{ij}= \left\{
        \begin{aligned}
           &+\frac{2\pi}{\phi_0} B \cdot i_y, \leftarrow \text{hopping}\\
           &-\frac{2\pi}{\phi_0} B \cdot i_y, \rightarrow \text{hopping}\\
           &0,\uparrow,\downarrow \text{hopping}\\
           &+\frac{2\pi}{\phi_0} B\cdot L \cdot i_x, \uparrow \text{hopping}\\
           &-\frac{2\pi}{\phi_0} B\cdot L \cdot i_x, \downarrow \text{hopping}\\
        \end{aligned}  
     \right.
  \end{equation}
  For the next-nearest-neighbor hopping,
  \begin{equation}
     A_{ij}= \left\{
        \begin{aligned}
           &+\frac{2\pi}{\phi_0} B \cdot i_y, \swarrow \text{hopping}\\
           &-\frac{2\pi}{\phi_0} B \cdot i_y, \nearrow \text{hopping}\\
           &+\frac{2\pi}{\phi_0} B \cdot i_y, \nwarrow \text{hopping}\\
           &-\frac{2\pi}{\phi_0} B \cdot i_y, \searrow \text{hopping}\\
           &+\frac{2\pi}{\phi_0} B \cdot (L i_x-i_y), \swarrow \text{hopping(boundary crossing)}\\
           &-\frac{2\pi}{\phi_0} B \cdot (L i_x-i_y), \nearrow \text{hopping(boundary crossing)}\\
           &+\frac{2\pi}{\phi_0} B \cdot (L i_x+i_y), \nwarrow \text{hopping(boundary crossing)}\\
           &-\frac{2\pi}{\phi_0} B \cdot (L i_x+i_y), \searrow \text{hopping(boundary crossing)}\\
        \end{aligned}  
     \right.
  \end{equation}
  where $B=\frac{\Phi_0}{L^2}$ is the unit magnetic flux, and $i_x, i_y$ are the indices of site range between $1$ and $L$ in the $x$ and $y$ lattice directions. Various arrows represent the direction of hopping terms from site $i$ to $j$. Note that when $L \rightarrow \infty$, the magnetic field approaches  0, and the Hamiltonian goes back to the original one.  
  The method of adding z-{\text{f}}lux significantly reduces finite size effects. However, the magnetic field breaks the translation symmetry, i.e., the momentum $k$ is not valid for fermion. In the DQMC simulation, we add the z-{\text{f}}lux when measuring bosonic observables, e.g., bosonic susceptibility. While for fermionic observables e.g. spectral functions, superfluid density, we drop it.
  
  \begin{table}[h]
     \caption{\bf Various explored pairing channels.}
     \label{table1}
     \centering
     \begin{tabular}{|c|c|c|}
     \hline
     \textbf{Channel} & \textbf{Description} & \textbf{Definition} \\
     \hline
     $C_{\text{os,is}}$ &  On-site, $s$-wave, Intra-layer, spin-singlet & $\frac{1}{\sqrt{2}} (\hat c_{i1\uparrow} \hat c_{i1\downarrow}+ \hat c_{i1\downarrow} \hat c_{i1\uparrow})$ \\
     \hline
     $C_{\text{os,ts}}$ & On-site, $s$-wave, layer-triplet, spin-singlet & $\frac{1}{\sqrt{2}} (\hat c_{i1\uparrow} \hat c_{i2\downarrow}- \hat c_{i1\downarrow} \hat c_{i2\uparrow})$ \\
     \hline
     $C_{\text{os,st0}}$ & On-site, $s$-wave, layer-singlet, spin-triplet($S=0$) & $\frac{1}{\sqrt{2}} (\hat c_{i1\uparrow} \hat c_{i2\downarrow}+ \hat c_{i1\downarrow} \hat c_{i2\uparrow})$ \\
     \hline
     $C_{\text{os,st1}}$ & On-site, $s$-wave, layer-singlet, spin-triplet($S=1$) & $\frac{1}{\sqrt{2}} (\hat c_{i1\uparrow} \hat c_{i2\uparrow}- \hat c_{i2\uparrow} \hat c_{i1\uparrow})$ \\ 
     \hline
     $C_{\text{ns,is}}$ & Nearest-neighbor, $s$-wave, Intra-layer, spin-singlet & $\frac{1}{\sqrt{8}} \sum_l f_{ns}(\delta_l) (\hat c_{i1\uparrow} \hat c_{i+\delta_l 1\downarrow}+ \hat c_{i1\downarrow} \hat c_{i+\delta_l 1\uparrow})$ \\
     \hline
     $C_{\text{ns,ts}}$ & Nearest-neighbor, $s$-wave, layer-triplet, spin-singlet & $\frac{1}{\sqrt{8}} \sum_l f_{ns}(\delta_l) (\hat c_{i1\uparrow} \hat c_{i+\delta_l 2\downarrow}- \hat c_{i1\downarrow} \hat c_{i+\delta_l 2\uparrow})$ \\
     \hline
     $C_{\text{ns,st0}}$ & Nearest-neighbor, $s$-wave, layer-singlet, spin-triplet($S=0$) & $\frac{1}{\sqrt{8}} \sum_l f_{ns}(\delta_l) (\hat c_{i1\uparrow} \hat c_{i+\delta_l 2\downarrow}+ \hat c_{i1\downarrow} \hat c_{i+\delta_l 2\uparrow})$ \\
     \hline
     $C_{\text{ns,st1}}$ & Nearest-neighbor, $s$-wave, layer-singlet, spin-triplet($S=1$) & $\frac{1}{\sqrt{8}} \sum_l f_{ns}(\delta_l) (\hat c_{i1\uparrow} \hat c_{i+\delta_l 2\uparrow} - \hat c_{i1\uparrow} \hat c_{i+\delta_l 2\uparrow})$ \\
     \hline
     $C_{\text{np,it0}}$ & Nearest-neighbor $p_x$-wave, Intra-layer, spin-triplet($S=0$) & $\frac{1}{\sqrt{4}} \sum_l f_{np}(\delta_l) (\hat c_{i1\uparrow} \hat c_{i+\delta_l 1\downarrow}+ \hat c_{i1\downarrow} \hat c_{i+\delta_l 1\uparrow})$ \\
     \hline
     $C_{\text{np,it1}}$ & Nearest-neighbor $p_x$-wave, Intra-layer, spin-triplet($S=1$) & $\frac{1}{\sqrt{2}} \sum_l f_{np}(\delta_l) \hat c_{i1\uparrow} \hat c_{i+\delta_l 1\uparrow}$ \\
     \hline
     \end{tabular}
  \end{table}
  \end{widetext}
  
  %\linenumbers
  
  \vspace{\baselineskip}
  \noindent {\bf Supplementary Note 2: Physical observables.}\\
  \vskip1mm
  \label{sec:seciib}
  
  In order to obtain the phase diagram of our model in the main text, we measure different physical observables in DQMC simulations, and analyze their behavior. Besides the main results presented in the main text, here we give a detailed description of the rest of them.
  
  \vskip1mm
  \noindent{\bf A. Pairing susceptibility and superfluid density.}
  \label{sec:seciib1}
  
  Superconductivity is expected to be enhanced near the QCP~\cite{Wang2013,Metlitski2015,Samuel2017}, but the detailed competition of the pseudogap, nFL and superconductivity in our system still needs to be revealed with different physical observables.
  
  \begin{figure}[tbp]
    \includegraphics[width=\columnwidth]{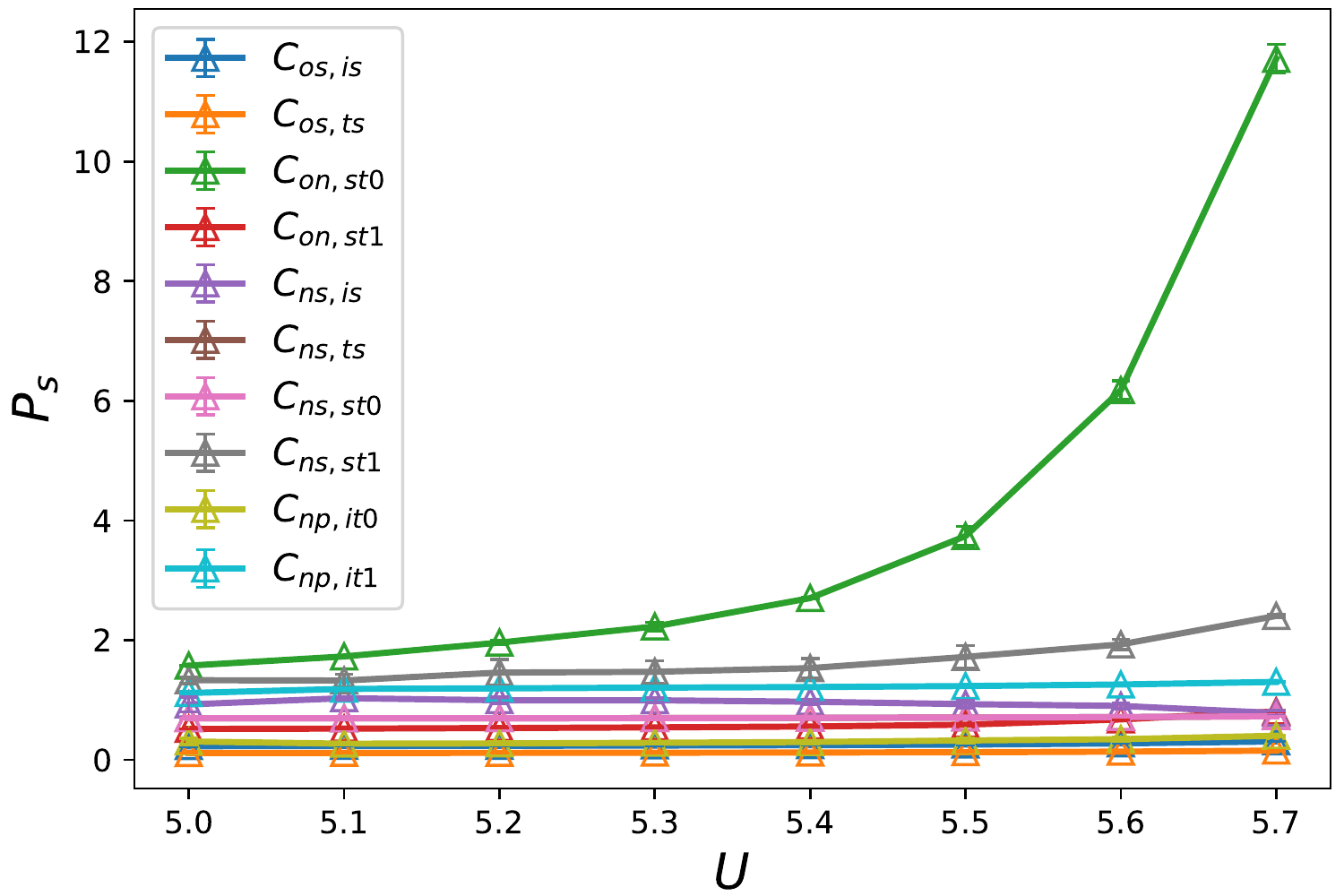}
    \caption{{\bf Comparision of different pairing channels versus $U$.} Pairing susceptibility of various pairing channels versus $U$ of $\beta=16$ of system size $L=12$. The green line corresponds to the orbital-singlet, spin-triplet channel, which is the putative dominant channel. When $U$ gets larger and approaches superconducting region, only $C_{\text{os,st0}}$ enhances remarkablely.}
    \label{fig:figRR1}
  \end{figure}
   
  As for the pairing, considering that the interaction is on-site and layer and spin symmetric, we compute a number of different on-site channels and find the strongest one occurs in the $s$-wave channel with orbital-singlet and spin-triplet $C_{\text{os,st0}}$, with the order parameter
  \begin{equation}
     \Delta(\mathbf{r})= C_{\text{os,st0}} = \frac{1}{\sqrt{2}} (\hat c_{\mathbf{r}1\uparrow} \hat c_{\mathbf{r}2\downarrow}+ \hat c_{\mathbf{r}1\downarrow} \hat c_{\mathbf{r}2\uparrow})
  \end{equation} 
  where 1,2 are layer indices. The detailed definition of the various pairing channels is listed in Supplementary Table ~\ref{table1}. Note $\hat x, \hat y$ represent unit vector along positive $x,y$ direction. $f_{\text{ns,is}}(\delta_l)=1$, for $\delta_l=\hat x, \hat y$ and $-1$, for $\delta_l=-\hat x, -\hat y$. $f_{\text{np,is}}(\delta_l)=1$, for $\delta_l=\hat x$, and $-1$ for $\delta_l=-\hat x$.
  
  We construct the pairing susceptibility defined as,
  \begin{equation}
     P_\text{s}=\frac{1}{L^2}\int_0^{\beta} \sum_i (\Delta^{\dagger}(\mathbf{r}_i,\tau) \Delta(\mathbf{0},0)).
  \end{equation}
  %where $\Delta(\mathbf{r},\tau)=\frac{1}{\sqrt{2}} (\hat c_{\mathbf{r}1\uparrow}(\tau) \hat c_{\mathbf{r}2\downarrow}(\tau)+ \hat c_{\mathbf{r}1\downarrow}(\tau) \hat c_{\mathbf{r}2\uparrow}(\tau))$. 
  $P_\text{s}$ captures the dynamic pair-pair correlation, which increases as temperature goes down. We show the results for all pairing channels and find no response therein in Supplementary Figure ~\ref{fig:figRR1} and Figure ~\ref{fig:figRR2} for other channels, except $C_{\text{os,st0}}$. Since the SC pair has quasi-long-range order below $T_\text{c} (T_{\text{KT}})$, $P_\text{s}$ will exhibit scaling behavior with system size as $P_\text{s} = L^{2-\eta} f(L\cdot \exp(-\frac{A}{(T-T_\text{c})^{1/2}}))$ as $T$ approches $T_\text{c}$ from above, and thus at $P_\text{s} \propto L^{2-\eta}$ with $\eta_{\text{KT}}=1/4$ at the transition at the thermodynamic limit~\cite{Paiva2004,costa2018phonon,ChuangChen2021}. We show the evolution pairing susceptibility of various pairing channels approaching QCP in Supplementary Figure ~\ref{fig:figRR1} and Figure ~\ref{fig:figRR2}, from which one reads only $s$-wave orbital-singlet and spin-triplet channel plays the crucial role of superconductivity.
  
  \begin{figure}[tbp]
    \includegraphics[width=\columnwidth]{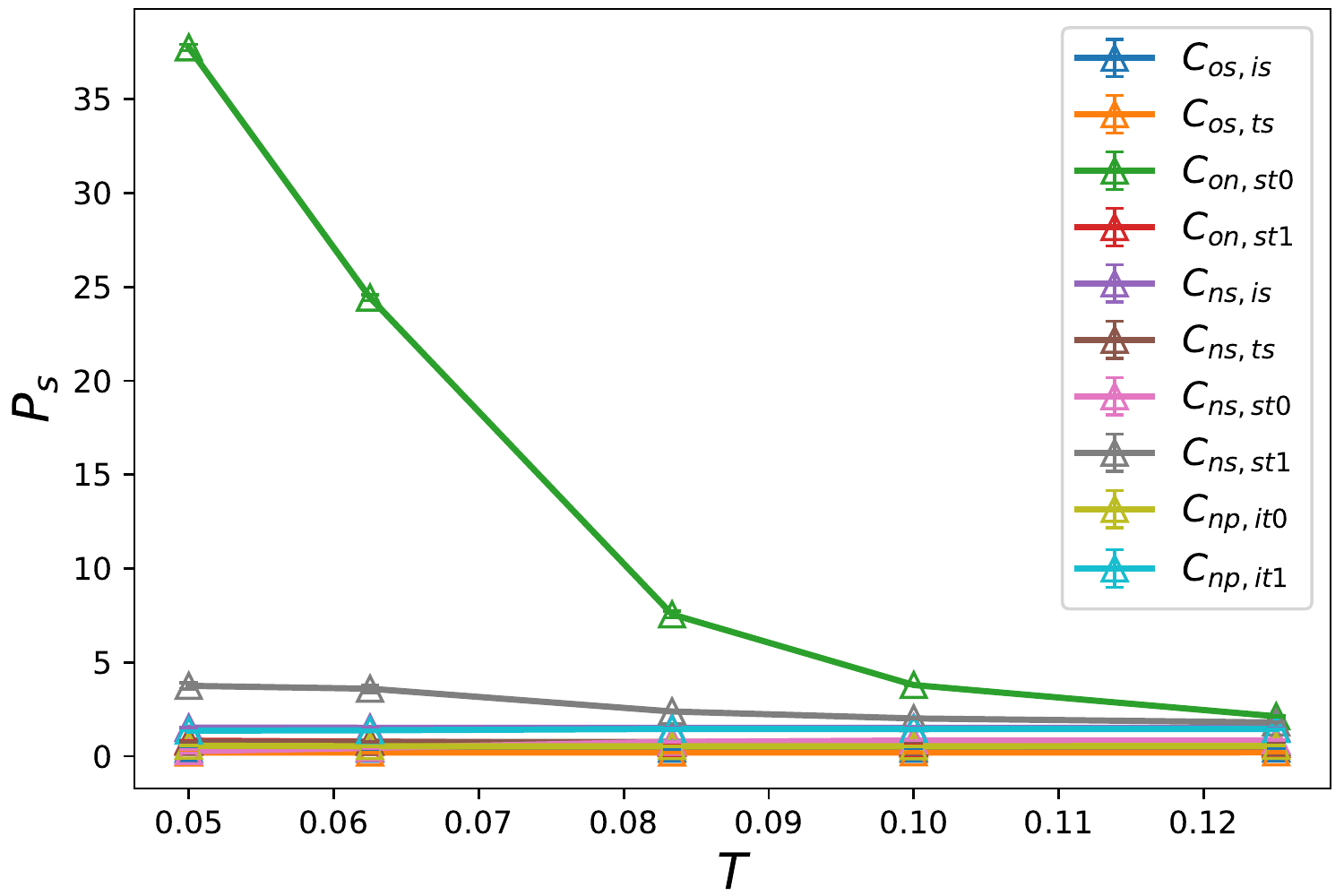}
    \caption{{\bf Comparision of different pairing channels  versus $T$.} Pairing susceptibility various pairing channels versus $T$ of $U=6.0$ of system size $L=12$. The green line corresponds to the orbital-singlet, spin-triplet channel, which is the putative dominant channel. When $T$ gets lower and enters superconducting region, only $C_{\text{os,st0}}$ enhances remarkablely.}
    \label{fig:figRR2}
  \end{figure}
  
  Further supporting evidence for the establishment of quasi-long-range order of $s$-wave pairing is the superfluid density $\rho_\text{s}$. $\rho_\text{s}$ describes tendency towards pairing, and is regarded as a measure of the ratio of superconducting electron density over compared to all the itinerant electrons~\cite{Scalapino1992,Scalapino1993}. The method to calculate $\rho_\text{s}$ is derived from the linear response to an external magnetic field. Adding a vector potential $A_x(\mathbf{r}_i,t)$ with harmonic frequency $\omega$ to the bond of the free fermion system and expanding to second order, one can deduce that the total induced current density $J_x(\mathbf{q},\omega)$ is,
  \begin{equation}
     \langle J_x(\mathbf{q},\omega) \rangle = -\left[ \langle -k_x \rangle -\Lambda_{xx}(\mathbf{q},\omega) \right] A_x(\mathbf{q},\omega)
  \end{equation}
  where $k_x$ is the kinetic energy density. $\Lambda_{xx}(\mathbf{q},\omega)$ is the current-current correlation function, which is associated with the paramagnetic current density $j_x^{\sigma}(\mathbf{r}_i,\tau)$, 
  \begin{equation}
     \Lambda_{xx}(\mathbf{q})=\frac{1}{4} \sum_{i\sigma\sigma'} \int_0^{\beta} d\tau e^{-i \mathbf{q} \mathbf{r}_i} \langle j_x^{\sigma}(\mathbf{r}_i,\tau) j_x^{\sigma'}(0,0) \rangle
  \end{equation} 
  where  $j_x^{\sigma}(\mathbf{r}_i,\tau)$ is defined as, 
  %\begin{equation}
  %   \begin{aligned}
  %   j_x^{\sigma}(\mathbf{r}_i,\tau) = it &\sum_{\lambda} ( \hat{c}^{\dagger}_{i,\lambda \sigma}(\tau) \hat{c}_{i+\hat{x},\lambda \sigma}(\tau) \\
  %   &- \hat{c}^{\dagger}_{i+\hat{x},\lambda \sigma}(\tau) \hat{c}_{i,\lambda \sigma}(\tau) )
  %   \end{aligned}
  %\end{equation}
  \begin{equation}
     \begin{aligned}
        &j_x^{\sigma}(\mathbf{r}_i,\tau) \\
        &= it_1 \sum_{\lambda,\sigma} \left( \hat{c}^{\dagger}_{i,\lambda \sigma}(\tau) \hat{c}_{i+\hat{x},\lambda \sigma}(\tau) - \hat{c}^{\dagger}_{i,\lambda \sigma}(\tau) \hat{c}_{i-\hat{x},\lambda \sigma}(\tau) \right)\\
        &+ it_2 \sum_{\lambda,\sigma} \left( \hat{c}^{\dagger}_{i,\lambda \sigma}(\tau) \hat{c}_{i+\hat{x}+\hat{y},\lambda \sigma}(\tau) - \hat{c}^{\dagger}_{i,\lambda \sigma}(\tau) \hat{c}_{i-\hat{x}+\hat{y},\lambda \sigma}(\tau) \right) \\ 
        &+ it_2 \sum_{\lambda,\sigma} \left( \hat{c}^{\dagger}_{i,\lambda \sigma}(\tau) \hat{c}_{i+\hat{x}-\hat{y},\lambda \sigma}(\tau) - \hat{c}^{\dagger}_{i,\lambda \sigma}(\tau) \hat{c}_{i-\hat{x}-\hat{y},\lambda \sigma}(\tau) \right) 
     \end{aligned}
  \end{equation}
  The criteria of superconductivity comes from the Meissner effect that if the current density response of a superconductor in a static, $\omega=0$, long wavelength $q_y=0$, the London equation is given by,
  \begin{equation}
     J_x(q_y)=-\rho_{\text{s}} A_x(q_y)
  \end{equation}
  where, $\rho_{\text{s}}$ is the superfluid density to be calculated. The response is always transverse
  % $\mathbf{q} \cdot \mathbf{A}$
  such that one can take the different order of the long wavelength transverse and longitudinal limit and obtain,
  \begin{equation}
     \begin{aligned}
     \rho_\text{s}=\langle -k_x \rangle - \Lambda_{xx}(q_x=0, q_y \rightarrow 0, i\omega=0) \\
     0=\langle -k_x \rangle - \Lambda_{xx}(q_x \rightarrow 0, q_y =0, i\omega=0)
     \end{aligned}
  \end{equation}
  which means $\rho_\text{s}$ can be  calculated from the current-current correlation function. In the thermodynamic limit, one expects that $\rho_\text{s}$ has a universal jump at the transition point, which according to the renormalization theory of the BKT phase transition~\cite{Nelson1977} is $\rho_{\text{s}}=\frac{2T_{\rho}}{\pi}$. We thus obtain $T_{\rho}$ by plotting $\rho_{\text{s}}(T)$, and looking for the crossing of $\rho_{\text{s}}$ with $\frac{2T}{\pi}$. It is also noted that in the correlated electron systems, such crossing temperature actually tends to overestimate the $T_{\text{c}}$ compared with that obtained from the pairing susceptibility~\cite{Paiva2004,costa2018phonon}, we have also confirmed such behavior in our simulation.
  % and would therefore pay more attention to the $T_{\text{c}}$ from the finite size scaling of $P_s$.
  
  \vskip1mm
  \noindent{\bf B. Bosonic susceptibilities.}
  \label{sec:seciib2}
  
  In the DQMC, we compute the dynamic bosonic susceptibility
  \begin{equation}
     \chi\left(h,T,\mathbf{q},\omega_n \right)=\frac{1}{L^2} \int d\tau \sum_{ij} e^{i \omega_n \tau - i \mathbf{q} \mathbf{r}_{ij}}  \langle {\mathbf{\theta}_i}(\tau) {\mathbf{\theta}_j}(0)\rangle.
  \end{equation}   
  For the bare rotor model, the behavior of dynamic susceptibility has the standard form,
  \begin{equation}
    \chi_0(\mathbf{q},\omega_n) = \frac{1}{\omega^2_n+q^2+\xi^{-2}_{\text{c}}},
    % \frac{1}{\omega^2_n-q^2-\xi^{-2}_{\text{c}}},
    % AK PLEASE CONFIRM CORRECTION
  \label{eq:eq26}
  \end{equation}
  where $\xi_{\text{c}}$ is the correlation length of bosonic field, which diverges at the critical point (in the lattice simulation such divergence is parameterized by the inverse distance towards the QCP in terms of the control parameter of the transition such as $U-U_{\text{c}}$). 
  When taking $q=0, \omega_n=0$, the dynamic susceptibility goes back to the uniform static susceptibility $\chi$, which only depends on temperature and tuning parameter $U$. Our results of the uniform static susceptibility of the coupled system can not be captured by generic the Curie-Weiss form in Supplementary Equation ~\eqref{eq:eq26}. Non-monotonous behaviour versus temperature is observed with fixing $U$ and reducing $T$ in a wide parameter range in the phase diagram (for example, as will be discussed in Supplementary Figure ~\ref{fig:figS2}). This is the interesting  fact that the coupling has altered the nature of the scaling in QCP and in our case such behavior serves as a signature of substantial superconducting fluctuations.
  
  As shown in the Fig.\ 4 in the main text, the $\omega_n$-dependence dynamic susceptibility follows a non-conserved bosonic order rule~\cite{Chubukov2014}, and displays continuous behavior at $\omega_n=0$, from which the novel quantum critical scaling behavior of our system is fully revealed. In the DQMC simulations, we also explore $q$-dependence and show the results in the following section.
  
  \vspace{\baselineskip}
  \noindent {\bf Supplementary Note 3: Pseudogap, superconductivity and the representative scans in the phase diagram.}\\
  \vskip1mm
  \noindent{\bf A. Scan at the maximum $T_{\text{c}}$.}
  
  We start with the scan at $U=6$ as a function of reducing temperature. In Fig.\ 2a in the main text, $N(\omega)$ is presented at various temperatures. It is clear that there emerges a pseudogap at the temperature $T=0.1 \ (\beta=10)$, and as the temperature is further reduced, the pseudogap steadily widens, and at the temperature of $T=0.05 \ (\beta=20)$, the full gap opens in the single-particle spectrum. 
  
  \begin{figure}[tbp]
    \includegraphics[width=\columnwidth]{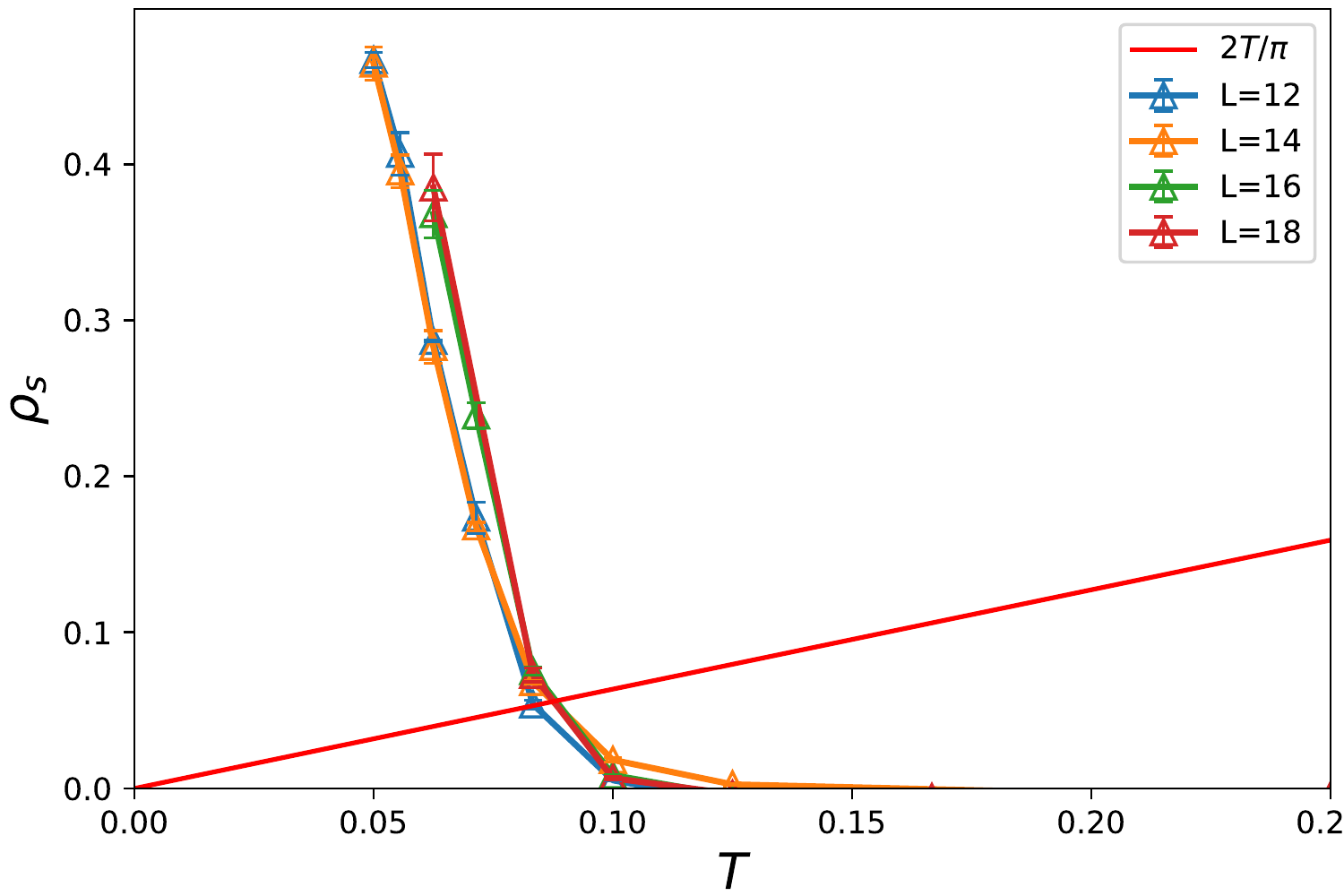}
    \caption{{\bf Superfluid density at the maximum $T_{\text{c}}$.} Superfluid density $\rho_{\text{s}}$ versus temperature at $U=6$ for system sizes $L=12,14,16,18$. The onset temperature of SC fluctuation $T_\rho$ is approximated by the crossover temperature for curve of $\rho_{\text{s}}(L \rightarrow \infty)$ and linear function with slope $\frac{2}{\pi}$. For $L=12,14$, such temperature is at the scale of $T\sim 0.1$, consistent with the onset of pseudogap in Fig.\ 2a in main text.}
    \label{fig:figS1}
  \end{figure}
  
  \begin{figure}[tbp]
     \includegraphics[width=\columnwidth]{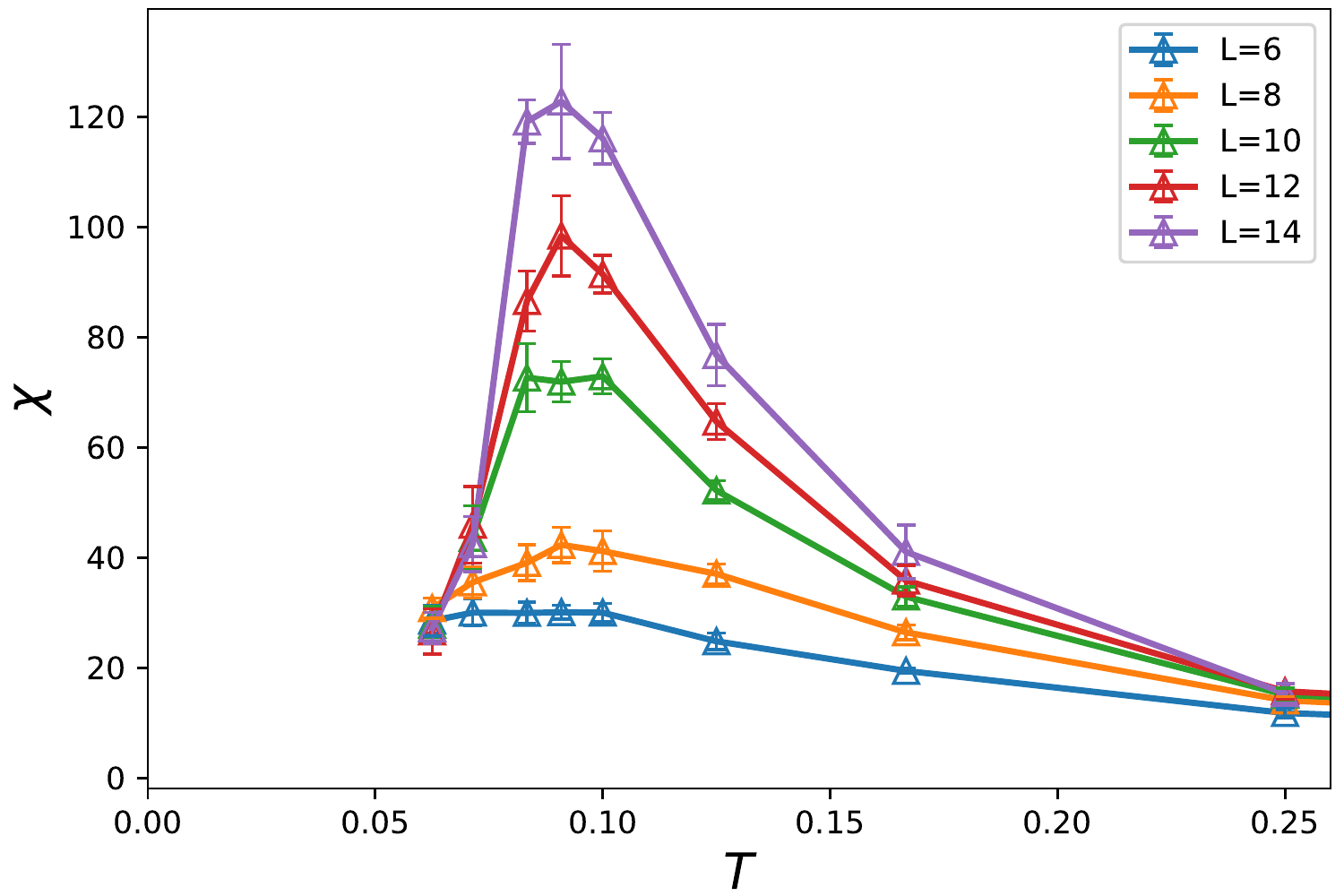}
    \caption{{\bf Bosonic static susceptibility at the maximum $T_{\text{c}}$.} Bosonic static susceptibility $\chi(\mathbf{q}=0,\omega=0)$ versus temperature at $U=6$ for system sizes $L=6,8,10,12,14$. Considering finite size effect, the nonmonotonous behavior of $\chi(T)$ depicts a crossover of two phase, whose transition temperature is approxiametly at $T=0.09$.}
    \label{fig:figS2}
  \end{figure}
  
  These two temperature/energy scales of the pseudogap region, are consistent with that in other physical observables. Supplementary Figure ~\ref{fig:figS1} shows  $\rho_{\text{s}}(T)$ along the same scan, and one sees that the crossing temperature for larger system sizes are at the onset of the pseudogap temperature $T=0.1$. Supplementary Figure ~\ref{fig:figS2} shows the uniform static susceptibility, and one observes a clear suppression of $\chi$, i.e. the deviation from the Curie-Weiss behavior is due to the onset of superconducting fluctuations, at the same temperature scale of $T=0.1$. As the pseudogap spectra gradually evolve into a full gap when temperature is decreasing, the superconducting fluctuation becomes stronger, and eventually renders the system into the quasi-long-range order of the $s$-wave pairing state. This can be seen in Fig.\ 2b in the main text, where the data collapse of the pairing susceptibility $P_{\text{s}}$ using KT phase transition critical characters for different system sizes are presented. One sees that at the temperature scale $T_{\text{c}} \approx 0.05$, a power-law divergence of $P_{\text{s}}$ is established. These criteria, together with the results from DOS and $\omega_n$-dependence bosonic dynamic susceptibility, reveals two distinct temperature/energy scales of the fermionic SC properties. 
  
  In addition, at $U=6$, the momentum dependence $\chi(|\mathbf{q}|,\omega=0)$, falls nicely with the power-law of $|\mathbf{q}|^{-2}$, as shown in Supplementary Figure ~\ref{fig:figS3}, which is similar to the bare rotor model in Supplementary Equation ~\eqref{eq:eq26}.
  
  \begin{figure}[tbp]
     \includegraphics[width=\columnwidth]{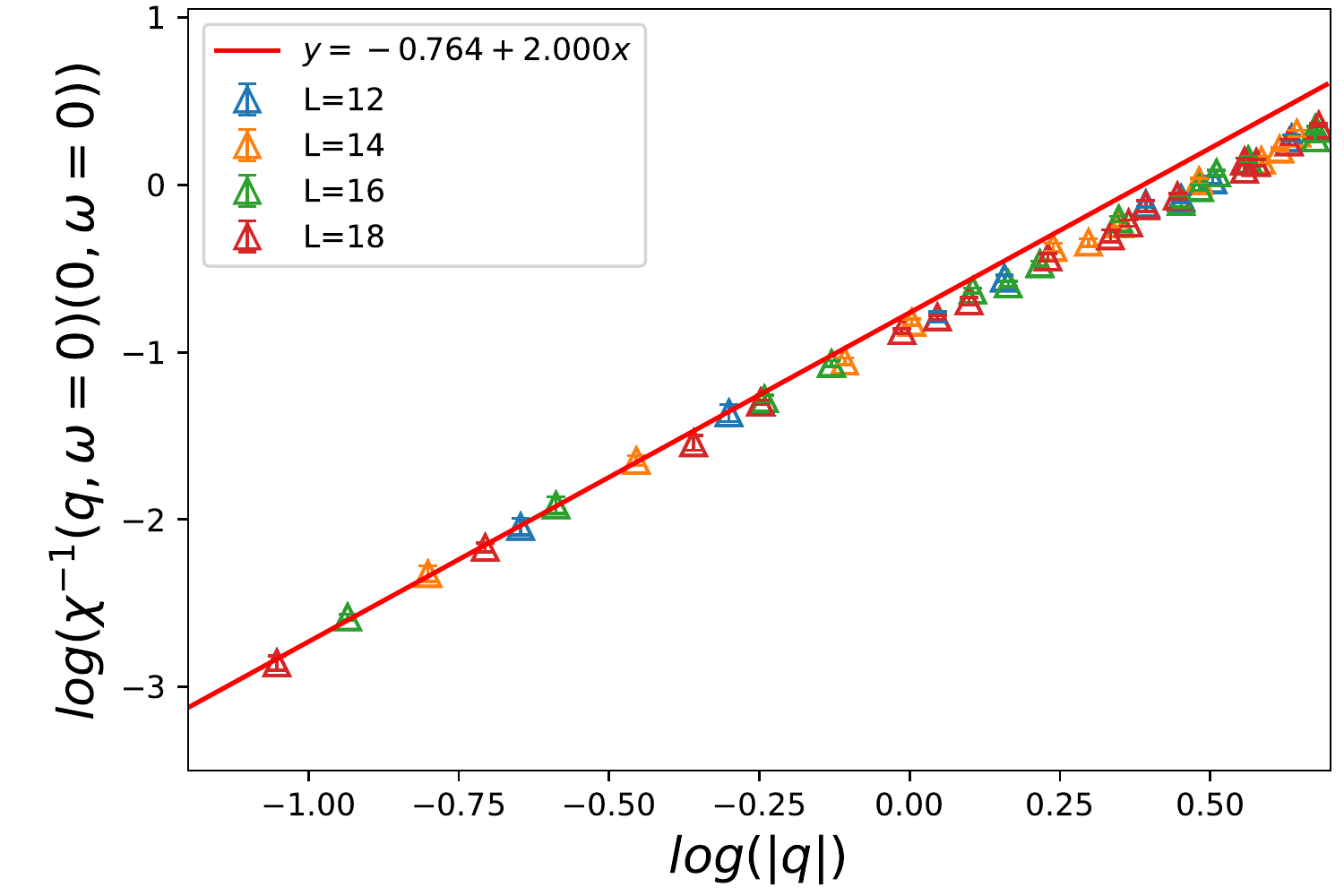}
    \caption{{\bf Dynamic bosonic susceptibility at the maximum $T_{\text{c}}$.} Dynamic bosonic susceptibility $\chi(|\mathbf{q}|,\omega=0)$ at $U=6$ with $T=0.1 \ (\beta=10)$ for system sizes $L=12,14,16,18$. The solid line is a guide to the eye of $\chi(|\mathbf{q}|)\sim 1/|\mathbf{q}|^2$. One sees that the power-law nicely captures the scaling behavior of data.}
    \label{fig:figS3}
  \end{figure}
  
  \begin{figure}[tbp]
    \includegraphics[width=\columnwidth]{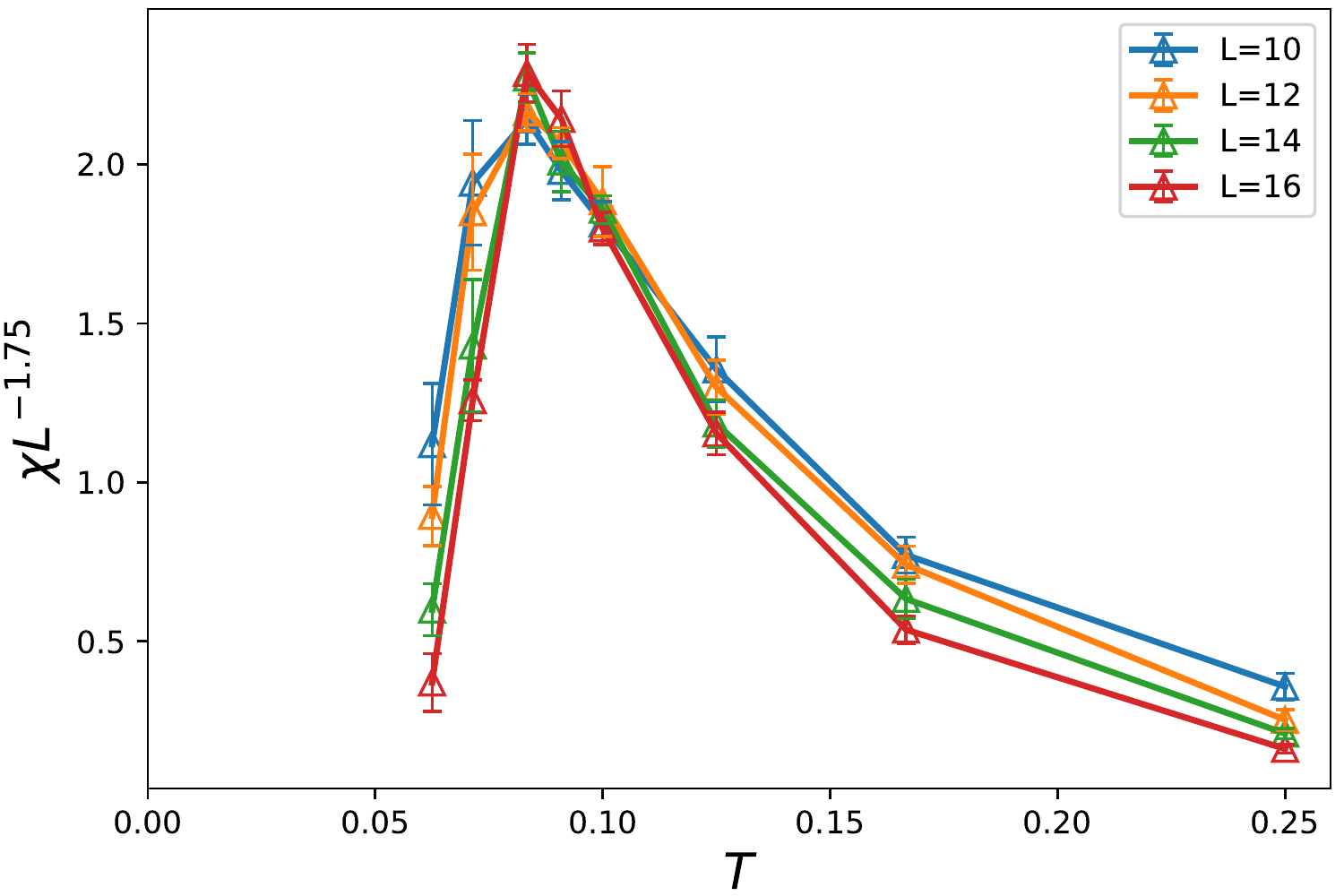}
    \caption{{\bf Rescaled bosonic static susceptibility at the re-entry regime.} Rescaled bosonic static susceptibility at $U=5.9$ versus temperature for system sizes $L=12,14,16,18$. Quasi-long-range order is expected to exist at the temperature where the rescaled susceptibility increases with system size, otherwise, in the disorder phase.}
    \label{fig:figS4}
  \end{figure}
  
  \vskip1mm
  \noindent{\bf B. Scans at the re-entry regime.}
  
  In this section, we focus at $U=5.9$, where the re-entry phenomenon is detected from the boundary of KT phase transition. We analyze the data in the KT phase at smaller $U$ versus $T$ near the QCP. We first use the scaling of the uniform bosonic susceptibility to determine the $U_{\text{KT}}$ as $\chi(U) = L^{2-\eta}f(L\cdot \exp(-\frac{A}{(U-U_{\text{KT}})^{1/2}})$ at fixed $T$ with $\eta=1/4$. It is expected that if one scales $\chi L^{-7/4}$, the curves of different system sizes will cross at the $U_{\text{KT}}$ and this is indeed what we saw in Supplementary Figure ~\ref{fig:figS4}. Here we fix $U=5.9$ and show the uniform susceptibility with different temperature. The KT scaling of the uniform susceptibility manifests, signifying the establish of the quasi-long-range order of the ferromagnetic rotor degrees of freedom. At $T \gtrsim 0.1$ and $T \lesssim 0.08$, the system is obviously located in the disordered phase. At intermediate temperatures at $\beta=11,12$ as calculated, there is clear evidence of forming quasi-long-range order.
  % Because the rescaled susceptibility of bigger size is above the small size  at fixed temperature, illustrating the fitting power of uniform susceptibility greater that critical exponent 1.75.
  % Please remove this sentence or explain it.
  Therefore, as a function of temperature, the system undergoes two KT phase transitions, i.e., shows the re-entry phenomenon.

  \begin{figure}[tbp]
     \includegraphics[width=\columnwidth]{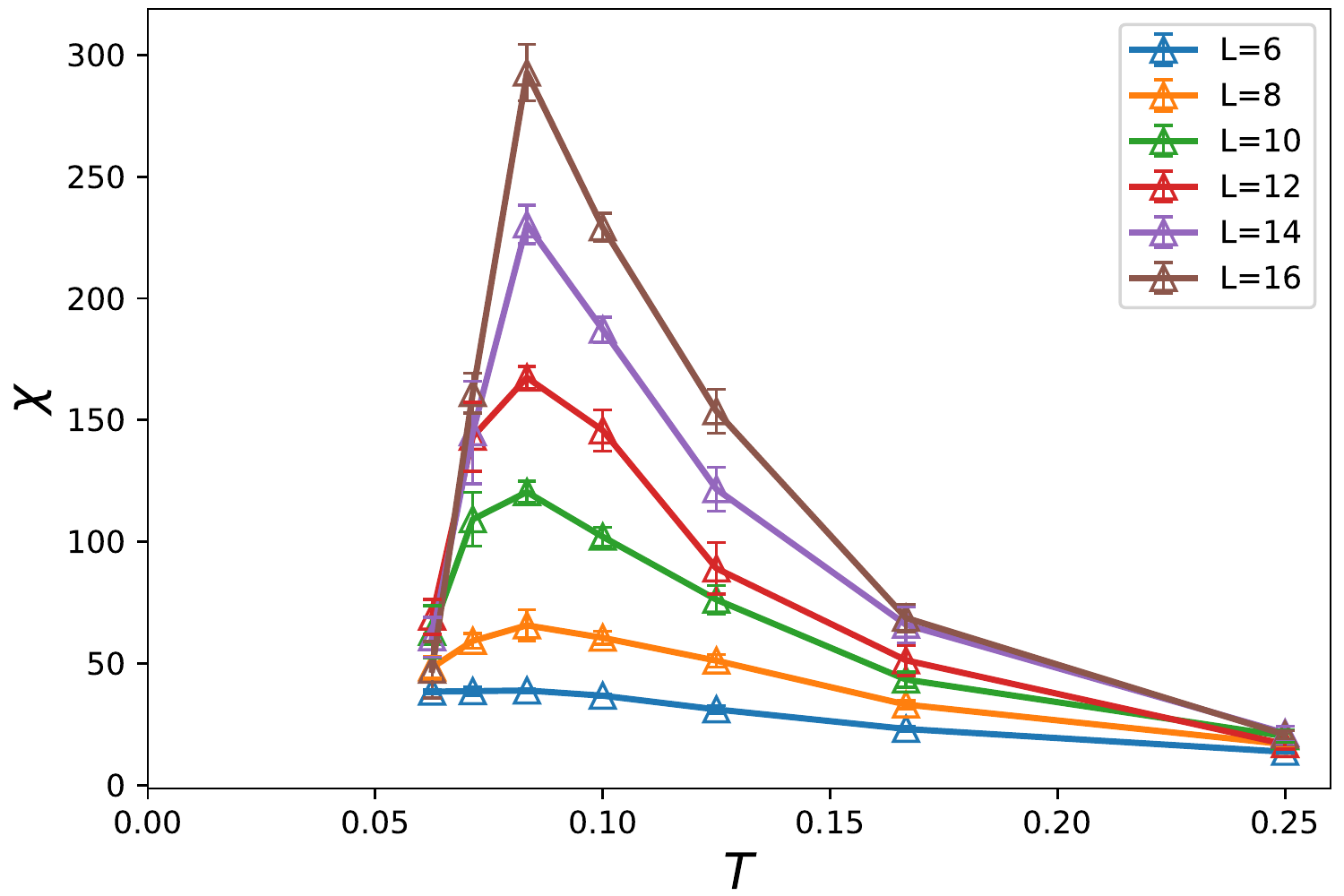}
    \caption{{\bf Bosonic static susceptibility at the re-entry regime.} Bosonic static susceptibility $\chi(\mathbf{q}=0,\omega=0)$ versus temperature at $U=5.9$ for system sizes $L=6,8,10,12,14$. Considering finite size effect, the nonmonotonous behavior of $\chi(T)$ depicts a crossover of two phase, whose transition temperature is approxiametly at $T=0.09$.}
    \label{fig:figS5}
  \end{figure}
  
  Furthermore, the uniform susceptibility also manifests the bending behavior along the temperature axis at $\beta=12$ shown in Supplementary Figure ~\ref{fig:figS5}. 
  
  %The $U_{\text{KT}}$ at other $T$ values are determined in the similar manner. Such phase boundary is consistent with that of the bare QRM model~\cite{WLJiang2019}.
  
  \vskip1mm
  \noindent{\bf C. Scans at ferromagnetic phase.}
  
\begin{figure}[tbp]
    \includegraphics[width=\columnwidth]{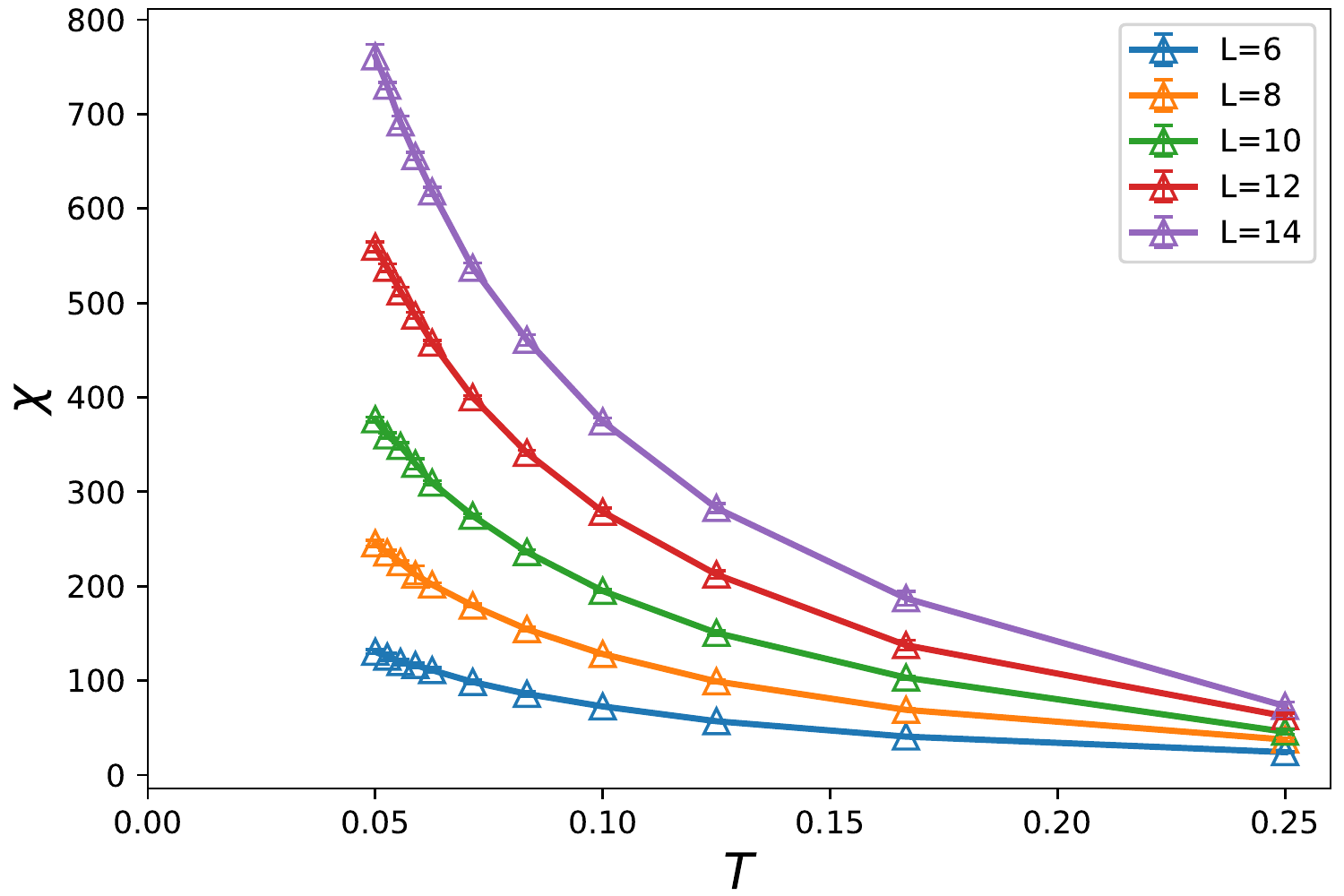}
    \caption{{\bf Bosonic static susceptibility at ferromagnetic phase.} Bosonic static susceptibility $\chi(\mathbf{q}=0,\omega=0)$ versus temperature at $U=5.5$ for system sizes $L=6,8,10,12,14$. The nonmonotonous behavior of $\chi(T)$ is not observerd at the lowest temperature $T=0.05$ calculated, indicating the pseudogap crossover temperature is lower than that.}
    \label{fig:figS6}
  \end{figure}

  When $U$ is decreased from the QCP to smaller values, the ferromagnetic properties of the bosonic part gradually increase. The pseudogap phase is strongly suppressed, and the crossover temperature drops to the temperature lower than we could explore. Similar to previous method, this can be seen from the bosonic static susceptibility. Supplementary Figure ~\ref{fig:figS6} shows that at $U=5.5$, the crossover temperature is lower than $T=0.05$. Thus we think the pseudogap phase disappears quickly at small $U$ in the ferromagnetic regime, as shown in the phase diagram in the main text.
  
  \vskip1mm
  \noindent{\bf D. Scans in the disordered phase.}
  
  The pseudogap phase found in the vicinity of QCP extends to the disordered phase at larger $U$. Here we present a temperature scan at $U=8$ with $N(\omega)$ at different temperatures. The behavior in Supplementary Figure ~\ref{fig:figS7} is similar with that in Fig.\ 2a in the main text, only that the onset of pseudogap now happens at slightly lower temperature of $T\sim 0.08$. However, the superconducting phase is clearly happening at a much lower temperature compared with that in QCP. For here even with $\beta=24$ the full gap is still not opened, in sharp constrast with the corresponding curve in phase diagram in the main text. Thus we think the SC phase domed at QCP does not extend as much as the pseudogap phase at large $U$, as shown in the phase diagram in the main text.
  
  \vskip1mm
  \noindent{\bf E. Scans for the diamagnetic fluctuations.}
  
  \begin{figure}[tbp]
     \includegraphics[width=\columnwidth]{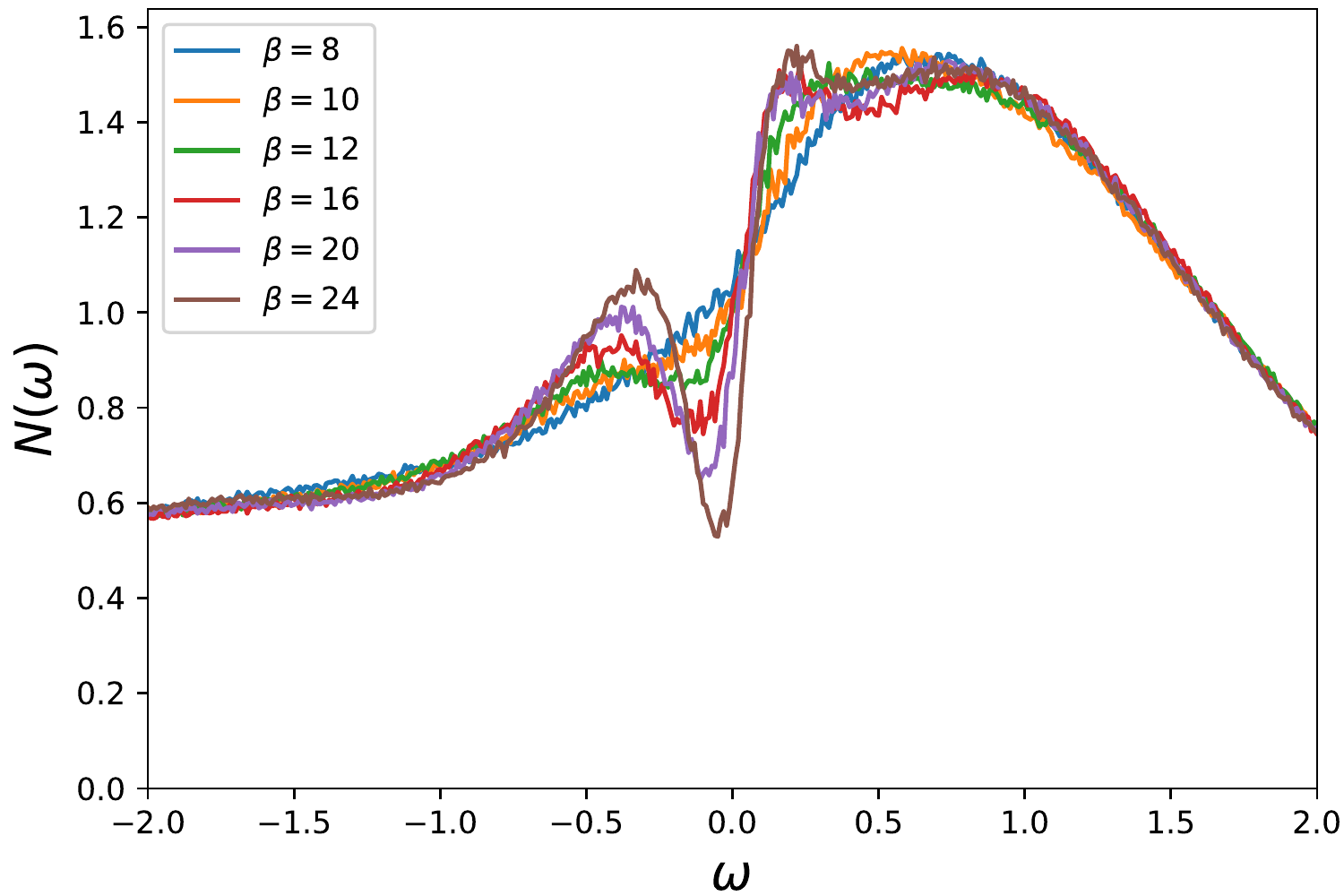}
    \caption{{\bf Local DOS in the disordered phase.} Local DOS $N(\omega)$ for various temperature at $U=8.0$ with $L=12$. The onset temperature of pseudogap phase is approxiametly at $T=0.08$. While at the lowest temperature at $T=0.04$, $N(\omega=0)$ is still far from 0, indicating the SC phase boundary is far less than $T=0.04$.}
    \label{fig:figS7}
  \end{figure}
  
  \begin{figure}[tbp]
     \includegraphics[width=\columnwidth]{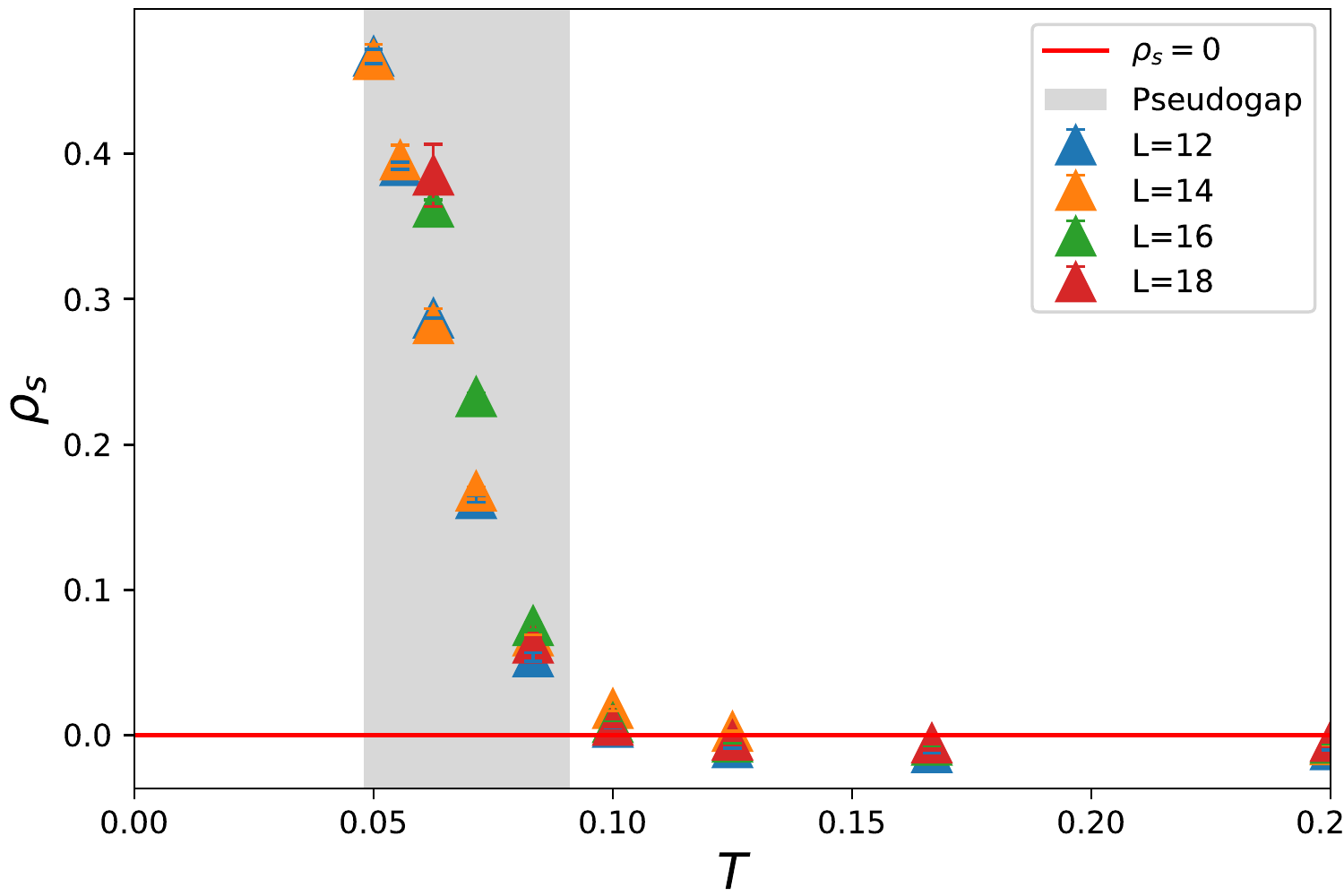}
    \caption{{\bf Superfluid density demonstrating diamagnetic fluctuation.} Superfluid density $\rho_{\text{s}}$ versus temperature at $U=6$ for system size $L=12,14,$ $16,18$. The grey area corresponds to the pseudogap region for fixing $U$. The red line denotes $\rho_{\text{s}}=0$, determining the crossover point for $T_{\text{dia}}$, where the finite size effect is negligible. $T_{\text{dia}} \approx 0.12$ for $U=6.0$.}
    \label{fig:figR1}
  \end{figure}
  
  Many experiments have observed the strong diamagnetic susceptibilities accompanying with pseudogap region. In the lattice model, the diamagnetic fluctuations corresponding to the Meissner effect are identified by measuring the superfluid density $\rho_{\text{s}}$~\cite{Schattner2016,Gerlach2017}. This quantity parametrizes the rigidity of the phase of the superconducting order parameter, which naturally leads to measuring the capacity of the superconductor to expel electromagnetic fields. We regard the value of $\rho_{\text{s}}$ goes from negative to positive as changing from paramagnetism to diamagnetism, and the curve consisting of such crossover points is identified as $T_{\text{dia}}$~\cite{Schattner2016,Gerlach2017}. Specifically, this behavior is observed by fixing the tuning parameter $U$, and scanning the temperature in Supplementary Figure ~\ref{fig:figS1}. Here, we put the same data set as that in Supplementary Figure ~\ref{fig:figR1}, but just reset the range of the y-axis to show the crossover with x-axis more clearly. The putative pseudogap region at $U=6$ is colored by grey, in which one can see the strong diamagnetic fluctuations, even extrapolating to infinite system size. The pseudogap region is totally under by the $T_{\text{dia}}$.
  
  \begin{figure}[tbp]
     \includegraphics[width=\columnwidth]{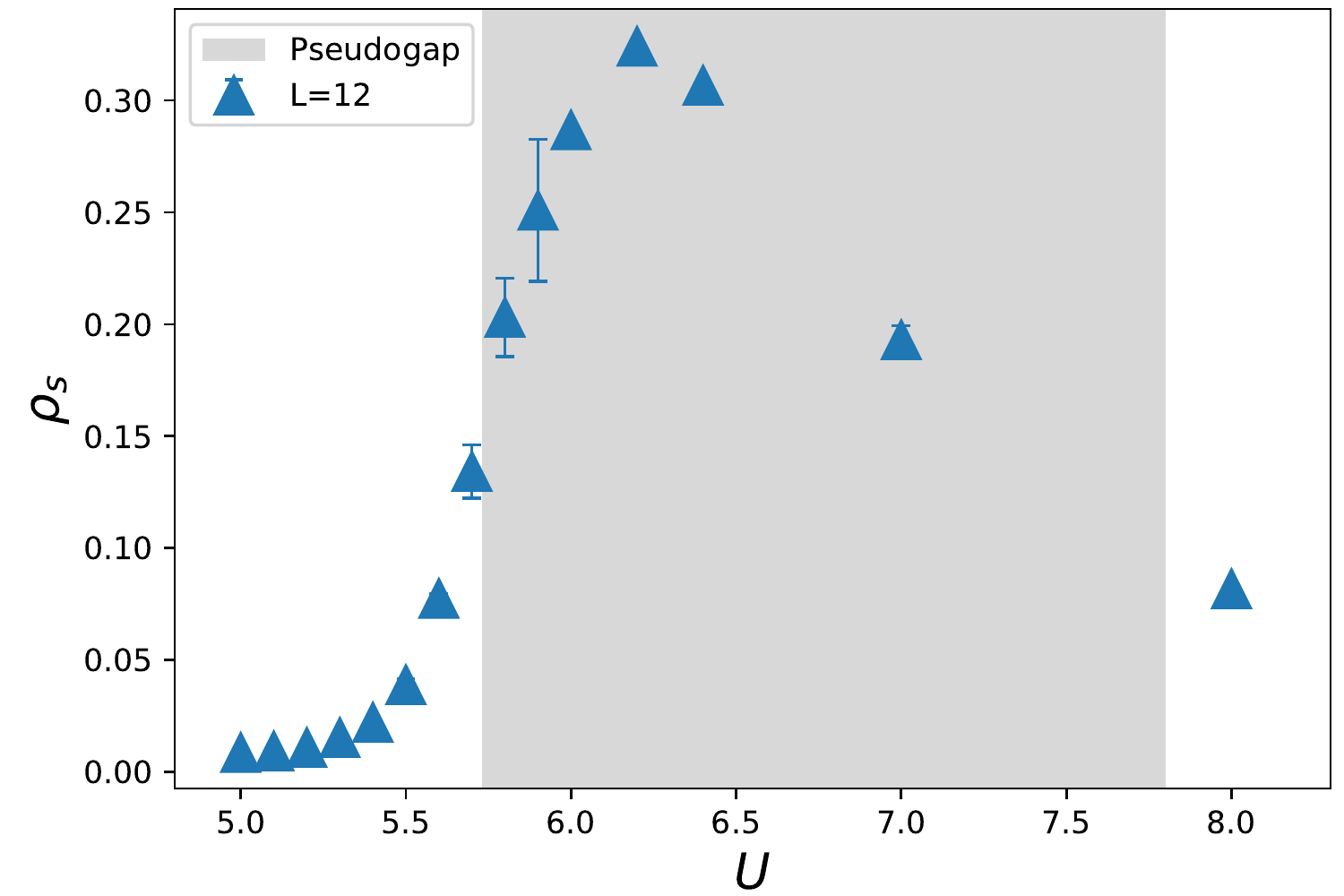}
    \caption{{\bf Superfluid density demonstrating diamagnetic fluctuation.} Superfluid density $\rho_{\text{s}}$ versus $U$ at inverse temperature $\beta=6$ for system size $L=8, 10, 12$. The grey area corresponds to the pseudogap region for fixing temperature. $\rho_{\text{s}}$ is significantly larger than 0 in the whole pseudogap region, denoting the strong diamagnetic fluctuation. }
    \label{fig:figR2}
  \end{figure}
  
  \begin{figure}[tbp]
     \includegraphics[width=0.46\textwidth]{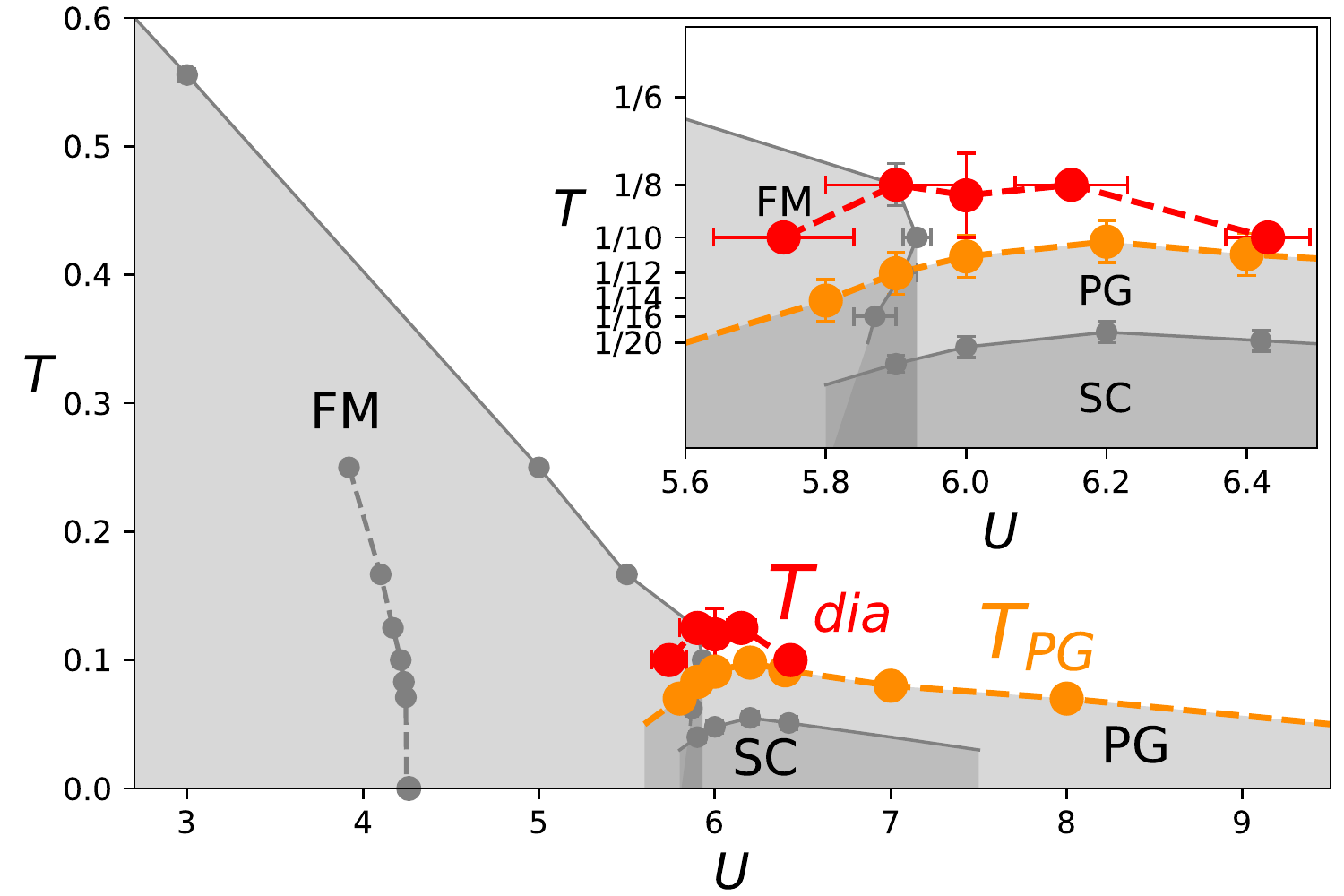}
     \caption{{\bf Boundary of the diamagnetic fluctuation in the phase diagram.} Phase diagram same as which in Fig.\ 1b in the main text. Other unconcerned regions and boundaries are colored by grey, except the boundary of $T_{\text{PG}}$, colored by orange. The boundary of $T_{\text{dia}}$ is shown by red. The points on the boundary is determined by the crossover method mentioned above, with fixed $U$ or temperature. The onset temperature of diamagnetism is obviously higher than the onset of pseudogap.}
     \label{fig:figR3}
  \end{figure}
  
  Besides, we fix the inverse temperature $\beta=16$, which is above the maximum of $T_{\text{c}}$, and draw $\rho_{\text{s}}$ versus $U$ in Supplementary Figure ~\ref{fig:figR2}. The putative pseudogap region is also colored by grey. One can see $\rho_{\text{s}}$ is large enough denoting the existing diamagnetic fluctuations before onset of true SC. Furthermore, we draw a boundary in the phase diagram representing the onset of diamagnetic fluctuations where $\rho_{\text{s}}$ changes sign, marked $T_{\text{dia}}$ in Supplementary Figure ~\ref{fig:figR3} below. The boundary of $T_{\text{dia}}$ covers the whole pseudogap region, and has the similar characters as $T_{\text{PG}}$.
  
  \vspace{1cm}
  
  \vspace{\baselineskip}
  \noindent {\bf Supplementary Note 4: Theoretical analysis.}\\
  \vskip1mm
  \noindent{\bf A. Modified Eliashberg theory.}
  
  We analyzed the QMC data for fermionic and bosonic response using the modified Eliashberg theory (mET), which is a low energy effective  dynamical theory for itinerant fermions near a QCP at finite temperatures.
   Within this theory, one obtains and solves the set of self-consistent equations for   fermionic self-energy $\Sigma (k, \omega_n) \approx \Sigma (\omega_m)$ and bosonic propagator $\Pi (q, \Omega_n)$, related to fermionic Green's function $G(k, \omega_n)$ and bosonic susceptibility $\chi (q, \Omega_n)$ as
  \begin{align}
    \label{eq:props}
    G^{-1}(\mathbf k, \omega_n) &= i\omega_n - \varepsilon_{\mathbf{k}} + i \Sigma(\mathbf k, \omega_n), \\
    \chi^{-1}(\mathbf q, \Omega_n) &= \chi_0^{-1}\left(r(T) + q^2 + c^2 \Omega_n^2 + \Pi(\mathbf q, \Omega_n) \right).
  \end{align}
  The input parameters for mET are 
  \begin{equation}
    \label{eq:params}
    k_{\text{F}},v_{\text{F}}, c, \chi(T)=\frac{\chi_0}{r(T)},\gb.
  \end{equation}
  These are, respectively, the fermionic Fermi vector, Fermi velocity, bosonic velocity, 
  static bosonic susceptibility, and the effective static boson-{\text{f}}ermion vertex. 
  At a QCP, 
  $r(T)\to 0$. 
  At the bare level, 
  $\gb = K^2 \chi_0$, see Eq.\ (1) in the main text, but
   it get substantially renormalized 
   by fermions with energies of order of the bandwidth, and we 
   treat  $\gb$ 
   as a fitting parameter. 
   The details of mET approach have been  
   discussed in Refs.~\cite{AKlein2020,XYXu2020}, and we refer an interested reader to those works for details.
   
   The fermionic self energy 
   consists of a thermal contribution $\Sigma_{\text{T}}(\omega_n)$ and quantum contribution $\Sigma_{\text{Q}}(\omega_n)$,
  \begin{equation}
    \label{eq:sig-def}
    \Sigma(\omega_n) = \Sigma_{\text{T}}(\omega_n)+ \Sigma_{\text{Q}}(\omega_n).
  \end{equation}
  The thermal contribution is a solution of a self-consistent equation,
  \newcommand{\w}{\omega}
  \begin{equation}
    \label{eq:sig-T-def}
    \Sigma_{\text{T}}(\w_n) = \frac{\gb T}{\pi}\frac{
      \mathcal{S}\left(\mathcal A_n\right),}{ |\w_n+\Sigma_{\text{T}}(\w_n)|}
  \end{equation}
  where 
  (for $n \geq 0$), 
   $\mathcal A_n = \frac{ v_{\text{F}}\sqrt{r(T)}}{|\w_n +\Sigma_{\text{T}}(\w_n)|}$ and $\mathcal{S}(x) = \frac{\cosh^{-1}(1/x)}{\sqrt{1-x^2}}$. 
   The quantum contribution is
  \begin{align}
    \Sigma_{\text{Q}}(\w_n) = \frac{
      \gb T }{\pi}\sum_{m\neq n}\frac{\mathcal{T}\left(\mathcal A_m, \mathcal B_{mn}\right)} {|\w_m+\Sigma_{\text{T}}(\w_m)|},
  \end{align}
  where $ \mathcal B_{mn} = \frac{(\gb k_{\text{F}}v_{\text{F}}|\omega_n-\omega_m|/\pi)^{1/3}}{|\w_m +\Sigma_{\text{T}}(\w_m)|}$ and
  \begin{align}
    \label{eq:T-def}
    \mathcal{T}(x,y) &= \int_0^{\infty} \frac{z^2 dz}{\sqrt{z^2+1}(z^3+ z x^2 + y^3)}.
  \end{align}
  The expressions are cumbersome, but 
  allows one to straightforwardly compute $\Sigma_{\text{T}}$ and $\Sigma_{\text{Q}}$ numerically.
  The outcome of the computations is the following. At $T\to 0$, $\Sigma_{\text{T}}$ vanishes and $\Sigma_{\text{Q}} \propto \omega_m^{2/3}$, leading to the well-known nFL fermionic behavior and $z=3$ dynamical scaling.
  At a finite $T$, there exists a wide range of temperatures, where  the 
  variations of $\Sigma_{\text{T}}$ and $\Sigma_{\text{Q}}$ with $\w_n$ 
  nearly compensate one another, leading to a fairly flat total self energy $\Sigma (\omega_m)$. 
  Roughly, this happens because $\Sigma_{\text{T}}$ ($\Sigma_{\text{Q}}$) are decreasing (increasing) functions of $\omega_n$. 
  
  The bosonic self-energy has the form,
  \newcommand{\W}{\Omega}
  \begin{align}
    \label{eq:Pi-sum-def}
    &\Pi(\mathbf q, \W_n) = \frac{2\gb T k_{\text{F}}}{v_{\text{F}}}\times\nonumber\\
    &~~\sum_{m=-n}^{-1} \frac{1}{\sqrt{(\W_n + |\Sigma(\w_{n+m})|+|\Sigma(\w_m)|)^2 + v_{\text{F}}^2 q^2}}
  \end{align}
  For $v_{\text{F}} q \gg \w_n,\Sigma$,  and at low $T$, $\Pi \propto \W_n/(v_{\text{F}} q)$ 
  has the form of a 
  canonical Landau damping of the bosons.  
  This gives rise to $z=3$ scaling. 
  However, 
  at  $q=0$ and at 
  a  small but finite 
  $\W_n$,
  \begin{equation}
    \label{eq:Pi-small}
    \Pi \propto \frac{\W_n}{\Sigma_{\text{T}}(\w_0)},
  \end{equation}
  and the scaling changes to $z=2$.
  
  \begin{figure*}[tbp]
    \centering
    \includegraphics[width=0.45\hsize]{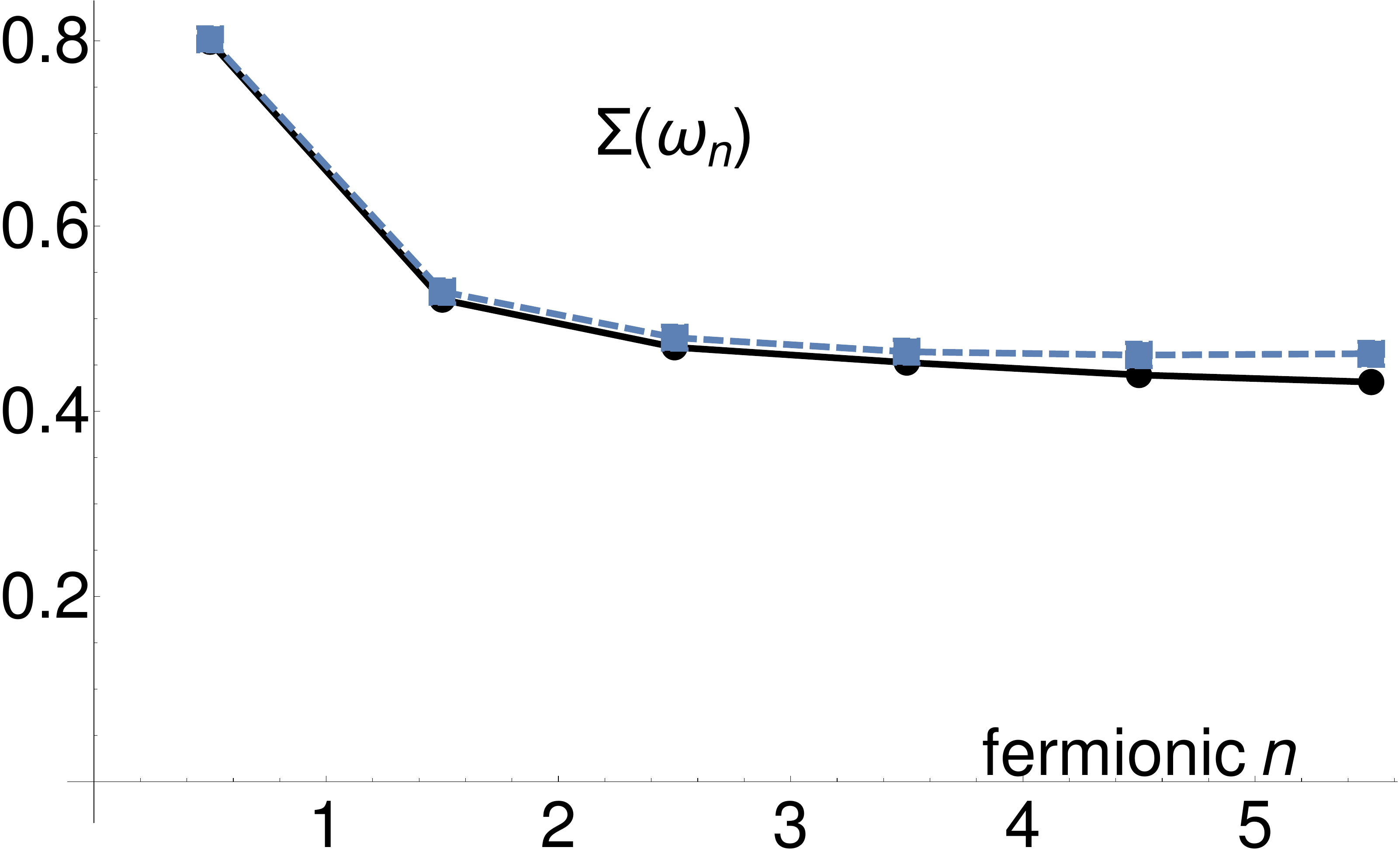}
    \includegraphics[width=0.45\hsize]{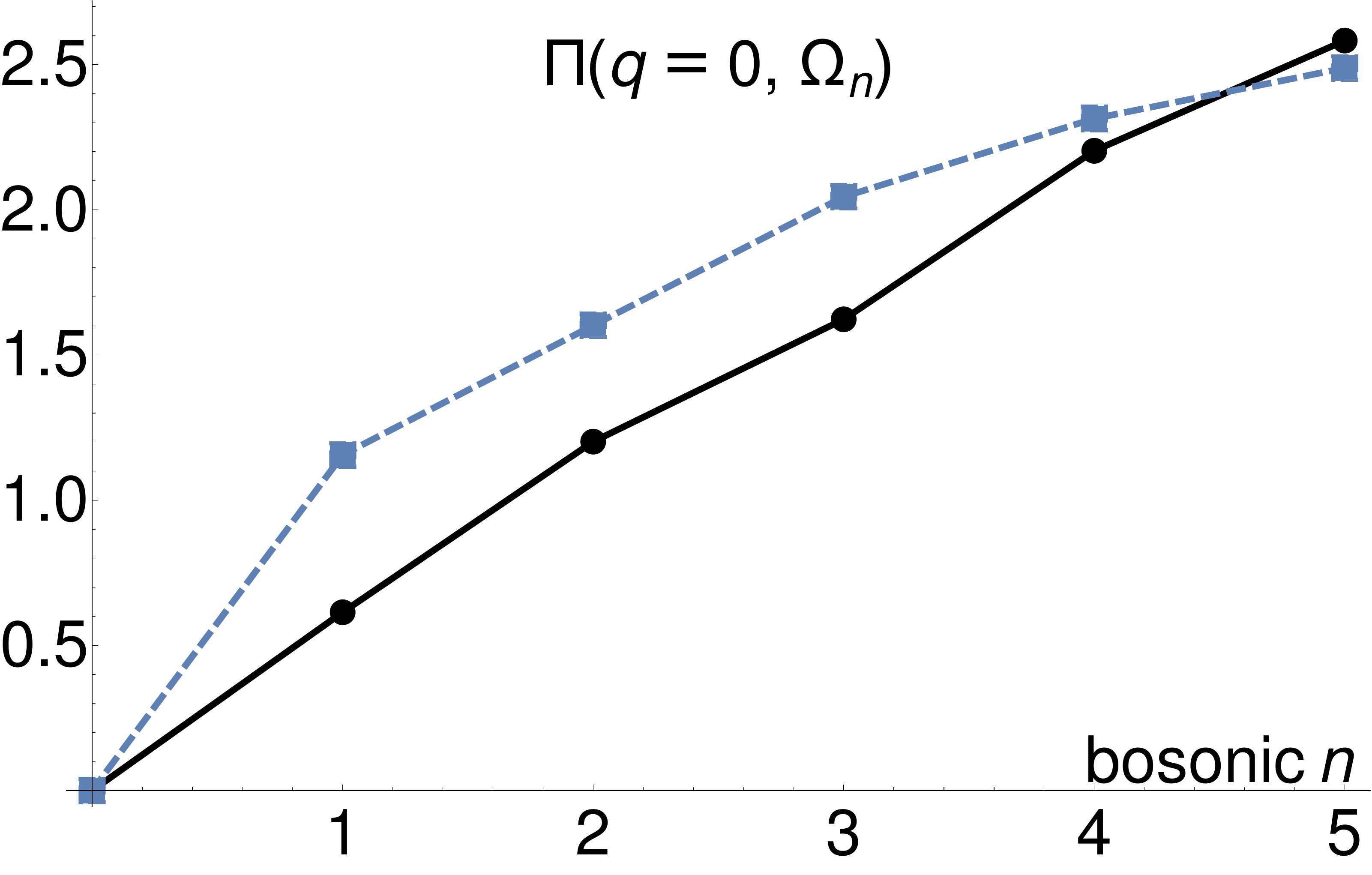}
    \caption{{\bf Comparison of QMC data with MET predictions.} (a) The fermionic self energy along the FS diagonal for $U=6, T=0.1$. The black circles are QMC data, and the blue squares are the theoretical prediction. (b) The bosonic self energy (for $\mathbf q = 0$).}
    \label{fig:QMC-MET}
  \end{figure*}
  
  Supplementary Equation \eqref{eq:Pi-sum-def} is justified for small/moderate frequencies. 
  At larger $\Omega_n$, vertex corrections play a role~\cite{Chubukov2005,Chubukov2014,Maslov2016}. 
  In the cases 
  when a
  boson represents a conserved quantity, 
  vertex corrections  remove 
  the dependence of $\Pi (q, \Omega)$ 
  on the self-energy, as required by the 
  Ward identity.
  However, in almost all QMC simulations to date, 
  the boson is not a conserved quantity, and therefore the effect of vertex corrections must be computed in case-by-case basis. 
  The results are that for an Ising spin, ladder vertex corrections are strong, and  the dressed 
  $\Pi (q, \Omega) \propto \W_n/\sqrt{\W_n^2+v_{\text{F}}^2q^2}$,  obeys an 
  ``effective'' Ward identity.
  Any violation of this identity must arise from additional diagrams, e.g. Aslamasov-Larkin diagrams, which  are expected to be weak. For an SU(2) spin, vertex corrections  are also strong, but do not
  cancel out the dependence of $\Pi$ on $\Sigma$.
  In this situation, 
  Eq. \eqref{eq:Pi-sum-def} is 
  good only for order of magnitude estimates
  for $\W_n \gtrsim v_{\text{F}} q$.  The case of an SO(2) spin,
  which we have in our simulations, is much better in this regard because 
  ladder vertex corrections actually vanish.
  In this situation,  corrections to Supplementary Equation \eqref{eq:Pi-sum-def} 
  only come from non-ladder diagrams. These  are normally quite small diagrams, so we expect Supplementary Equation \eqref{eq:Pi-sum-def}
  to be a fairly decent approximation.
  
  \vskip1mm
  \noindent{\bf B. Data Analysis.}
  
  Since MET is valid in the vicinity of a QCP, we picked data for $U=6$ to perform the analysis. This is because it is near the critical $U=5.9$, but the reentrance effect, shown in Fig.\ 1 in the main text, is weaker there. Nevertheless, the system develops both a PG and magnetic order for low $T$. This poses several constraints. First, we are limited to $T>0.1$ to avoid a significant pseudogap. This means that there are very few data points that are valid for a low-energy theory. We picked as a cutoff in frequency $\w_{\text{F}} = k_{\text{F}} v_{\text{F}} /2$, and as a cutoff in momentum $q_{\text{F}} = v_{\text{F}}^{-1}$, which leaves about 5 Matsubara frequencies within our window at $T=0.1$. Second, the presence of even a small Zeeman gap distorts the self energy such that the self-energy obeys
  \begin{equation}
    \label{eq:sig-QMC}
    \Sigma_{\text{FM}} = \Sigma(\w_n) + \frac{\Delta_{\text{FM}}^2}{\w_n}
  \end{equation}
  where $\Sigma(\w_n)$ appears in Supplementary Equation \eqref{eq:sig-def}. In terms of the inputs to our theory, Supplementary Equation \eqref{eq:params}, $k_{\text{F}}, v_{\text{F}}, \chi_0,r(T)$ were obtained from the band structure and from the QMC data for the bosonic propagator. Unfortunately, we were not able to extract $c$ reliably from the data, and so in our calculations we set $c=0$ for simplicity (we checked that varying $c$ doesn't qualitatively change our results). We then fit the fermionic self energy at the FS to Supplementary Equation \eqref{eq:sig-QMC}, using $\Delta_{\text{FM}}$ and $\gb$ as fitting parameters. We present the data for $\Sigma_{\text{FM}}$ along the BZ diagonal in Supplementary Figure \ref{fig:QMC-MET}(a), showing excellent agreement with the data. The fit parameters were
  \begin{equation}
    \label{eq:fits}
    \Delta_{\text{FM}} = 0.34 \pm 0.01, \gb = 6.3 \pm 0.2.
  \end{equation}
  $\Delta_{\text{FM}}$ is on order of an inverse lattice vector $\pi/L, L=12$, consistent with the splitting seen in e.g. Fig.\ 3(b) in the main text. $\gb$ is a bit higher than the bare $\gb_0 = 4.2$ obtained from the model parameters, and represents about a $20\% $ increase in the interaction vertex $K$, consistent with previous QMC at strong coupling. We expect the coupling in Supplementary Equation \eqref{eq:fits} to be somewhat over-estimated because we neglected $c>0$ effects. We checked our fits by comparing the theoretical and QMC bosonic self-energy. We present the comparison in Supplementary Figure \ref{fig:QMC-MET}(b), which shows a fairly good agreement. The quantitative discrepancies are not surprising, both because of the Zeeman gap and due to the issues discussed above and in the background section. For our purposes, it is enough that the theory correctly predicts the deviation from $z=3$ scaling, and that the slope of the theoretical and QMC data are comparable, confirming that our estimate of $\gb$ is reasonable.
  
  To confirm that the onset of the pseudogap in our simulations is consistent with theoretical predictions, we computed $T_{\text{PG}}$ within the $\gamma-$model (see Ref.~\cite{YMWu2020} and references within). To facilitate comparison with previous works, we supply here the conversion between the coupling $\gb$ in our model, and the effective coupling $\gb_\gamma$ in the $\gamma-$model. It is,
  \begin{equation}
    \label{eq:gb-gamma}
    \gb_\gamma = \frac{1}{2\pi^2}\frac{\gb^2}{k_{\text{F}} v_{\text{F}}}\left(\frac{2}{3\sqrt{3}}\right)^3. % Factors come from: 2 bosons, 2 fermions, a factor of 1/3 from momentum integration.
  \end{equation}

  Our model corresponds to a $\gamma=1/3$ model, for which $T_{\text{PG}} = 4.4 \gb_\gamma$. Using the extracted $\gb$ from Supplementary Equation \eqref{eq:fits} we find $T_{\text{PG}} = 0.08$, which is in good agreement with the measurement of $T_{\text{PG}} \sim 0.1$.

\vspace{\baselineskip}  
\bibliography{short}
\end{document}